\documentstyle[preprint,aps,eqsecnum,psfig]{revtex}
\tightenlines
\begin{document}

\draft

\title{From nonlinear scaling fields to critical amplitudes}

\author{Y. Meurice and S. Niermann\\ 
{\it Department of Physics and Astronomy, The University of Iowa, 
Iowa City, Iowa 52242, USA}}

\maketitle

\begin{abstract}

We propose to combine the nonlinear scaling fields 
associated with the 
high-temperature (HT) fixed point, with those associated with the unstable
fixed point, in order to calculate the susceptibility and other 
thermodynamic quantities.
The general strategy relies on 
simple linear relations between the HT scaling fields and the thermodynamic
quantities, and the
estimation of RG invariants formed out of the two sets of scaling fields.
This estimation requires convergent expansions in overlapping domains.
If, in addition, the initial values of the scaling fields 
associated with the unstable fixed point can be 
calculated from the temperature and the parameters appearing 
in the microscopic Hamiltonian, one can estimate the critical amplitudes.
This strategy has been developed using Dyson's hierarchical model 
where all the steps can be approximately implemented with good accuracy.
We show numerically that for this model (and a simplified version of it),
the required overlap apparently occurs, allowing an accurate determination of 
the critical amplitudes.

\end{abstract}

\vfill
\eject
\section{Introduction, motivations and main results}
It is well-known that the magnetic susceptibility  of a spin model
near its critical temperature can be parametrized  as
\begin{equation}
\chi= (\beta _c -\beta )^{-\gamma } (A_0 + A_1 
(\beta _c -\beta)^{
\Delta }+\dots )\ .
\label{eq:param}
\end{equation}
In this expression, the exponents $\gamma$ and $\Delta$ are 
universal and can be obtained from the calculation of the eigenvalues 
of the linearized renormalization group (RG) transformation. 
On the other hand, the critical amplitudes 
$A_0,\ A_1,\dots$ are functions of the microscopic
details of the theory. One can find universal relations \cite{privman} 
among these amplitudes  and the ones associated with 
other thermodynamic quantities,
however these relations do not 
fix completely the amplitudes. In the end, if we want a quantitative 
estimate of a particular amplitude, we need to perform a calculation 
which requires a knowledge of many details of the RG flows.
Such a calculation is in general a difficult, nonlinear, multivariable problem.
In this article we propose a general strategy
based on the construction of nonlinear scaling fields associated 
with {\it several} fixed points, to calculate the critical
amplitudes, and we demonstrate its feasibility in the case of Dyson's 
hierarchical model.

A common strategy in problems involving nonlinear flows 
near a singular point, is to construct a new 
system of coordinates for which the governing equations become 
linear. 
It seems intuitively clear that 
if the original problem is sufficiently nontrivial, 
normal form methods
can only work in some limited way, locally, because the flows of the 
nonlinear problem have global properties which do not 
match those of the linear flows. 
A well-known argument for the
inadequacy of such procedure (which extends beyond the special case of 
an expansion near a singular point) was provided 
by Poincar\'e \cite{poincare92} in the context of perturbed integrable
Hamiltonians. He discovered that even though it is 
possible to write a formal perturbative series for 
the action-angle variables,
some coefficients have ``small denominators'', 
and generically, the series are ill-defined. 
However, under some restrictions (formulated
according to some appropriate version of the K. A. M. theorem \cite{arnold88}),
perturbation theory can still provide interesting information.

Almost thirty years ago, Wegner\cite{wegner72}, introduced 
quantities that transformed multiplicatively under a RG 
transformation. He called them ``scaling fields'' and we will
use his terminology in the following.
Sometimes, one also uses the terminology ``nonlinear scaling field''
to distinguish them from the linear ones (see section \ref{sec:3steps} 
for details). In the following, ``scaling fields'' mean the nonlinear
ones and we will use the terminology ``linear scaling fields'' 
when necessary.
These fields play a central role in the presentation
of the basic ideas of the RG. 
They appear in almost any review on 
the subject (see for instance Ref. \cite{cardy96}).
As in the case of Hamiltonian dynamics, there
exists a formal series expansion for the scaling 
variables (see Eq. (4.9) in Ref. \cite{wegner72}). 
It is commonly assumed that the functions defined
with this procedure are analytic, at least within a
certain neighborhood of the fixed point.
However, for most non-trivial models,
it is very difficult to prove this assumption.
In particular, it is difficult to address the question of 
small denominators 
because it requires an accurate calculation of the eigenvalues of
the linearized RG transformation. 

If the small denominator problem can be controlled 
and if some {\it local} expansion is well-defined, there remain
several important {\it global} issues.
What is the domain of convergence
of this expansion?
How does the accuracy of an expansion with a 
finite  number of terms evolve when we move away from the 
fixed point?
Can different expansions have overlapping domain of convergence?
These important {\it global} issues are rarely discussed because of
practical limitations: in crossover regions, we need large
order expansions in many variables. Unfortunately, this 
problem has to be faced if we want to calculate all the critical 
amplitudes. 
In this article, we propose a general strategy to calculate directly the 
critical amplitudes. 
This strategy has been developed using Dyson's hierarchical model, 
where large order expansions in many variables are
practically feasible. All the numerical calculations presented hereafter 
were done with this model (or a simplified version of it).

The general point of view that we want to advocate here is that 
one should combine different sets of scaling fields.
Even though the scaling fields are almost always constructed 
in the vicinity of Wilson's fixed point, they can 
in principle be constructed near
any other fixed point. If one can find some overlap among 
the domains of convergence of these expansions
it is possible to reconstruct the flows, given their
initial values. 
In other words, we would like to develop a new analytical approach
to complement the existing methods
used to deal with the crossover between fixed points, namely, 
the Monte Carlo method
\cite{gonzalez87,luijten98,taro00}, a combination
of field-theoretical methods and mean field calculations 
\cite{bagnuls85,pelissetto98} or
the study of the entropy associated with the RG flows \cite{gaite96}.

\begin{figure}
\vskip125pt
\centerline{\psfig{figure=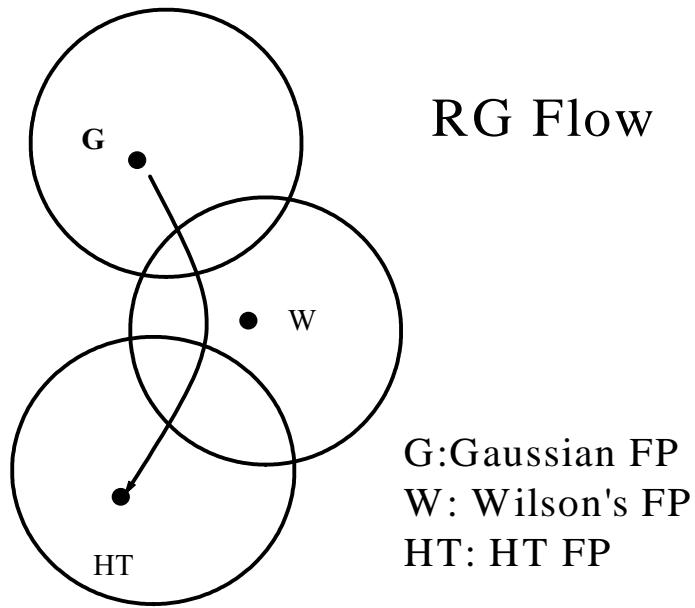,width=5in}}
\vskip125pt
\caption{Schematic representation of a RG flow starting near 
the Gaussian fixed point, passing near Wilson's fixed point
and ending at the stable high-temperature fixed point. The circles
represent the domains of validity of expansions of the scaling fields 
near the three fixed points.}
\label{fig:pic}
\end{figure}

In the following, we concentrate on the study of the RG 
flows in the symmetric phase of 
spin models having a nontrivial unstable fixed point.
Our goal is to calculate the critical amplitudes  
by constructing the scaling fields near the 
three relevant fixed points: the Gaussian fixed point (if relevant),
the unstable 
fixed point (sometimes called the IR fixed point or Wilson's fixed point), 
and the high-temperature (HT) fixed point. 
The idea is represented schematically in Fig. \ref{fig:pic}.

We propose to follow three specific steps to achieve this goal.
These steps correspond to a construction in  backward 
order, starting with the 
flows near the HT fixed point and ending with the initial conditions.
First, we express the thermodynamic quantities in terms of the 
scaling fields of the HT fixed point. Second, we use the scaling fields
of the unstable IR fixed point, to write the thermodynamic quantities
as a main singularity times a RG invariant quantity 
constructed out the two scaling fields. 
Third, we calculate the initial
values of the scaling fields associated with the unstable fixed point
in terms of the basic parameters appearing the microscopic Hamiltonian.
These three steps are explained in more detail in Section \ref{sec:3steps}.
This section is 
essential to understand the general ideas and the notations used later.
It should be noted that the first two steps are independent of the 
initial conditions and are in some sense universal. On the other hand,
the third step provides the initial data as a function of the basic
parameters such as the temperature or the ``bare parameters'' of 
Landau-Ginzburg models. Consequently, the method used to implement the
third step depends on the initial measure considered. For instance,
if we start with an initial measure near the Gaussian fixed point, 
perturbative field theoretical methods (Feynman diagrams) will be used.

Each of these three steps can in principle be 
implemented 
for spin models with nearest 
neighbor interactions in three dimensions, by using available expansions 
to describe the flows in each region. 
However, in order to discuss the behavior of the expansions in the 
crossover region, we need large order expansions.
Despite the existence of increasingly sophisticated methods used
for various expansions in nearest neighbor models 
(see e.g., Ref. \cite{campostrini01}), it is still 
a major time investment to perform these expansions.
In order to test the feasibility of the procedure, we have considered 
approximations for which large order expansions (see e.g., Refs. 
\cite{osc1,osc2}) can be reached 
more easily. Namely,
we have used
hierarchical approximations where only the local part 
of the measure is renormalized under a RG transformation. 
Well-known examples 
are the ``approximate recursion formula'' \cite{wilson71b} 
or Dyson's hierarchical model (HM)\cite{dyson69,baker72}.
Despite this approximation, the nonlinear aspects of the 
problem are still nontrivial. 
To fix the ideas, the calculations that we have 
performed required the determination of several thousands of coefficients 
in various expansions.

\begin{figure}

\centerline{\psfig{figure=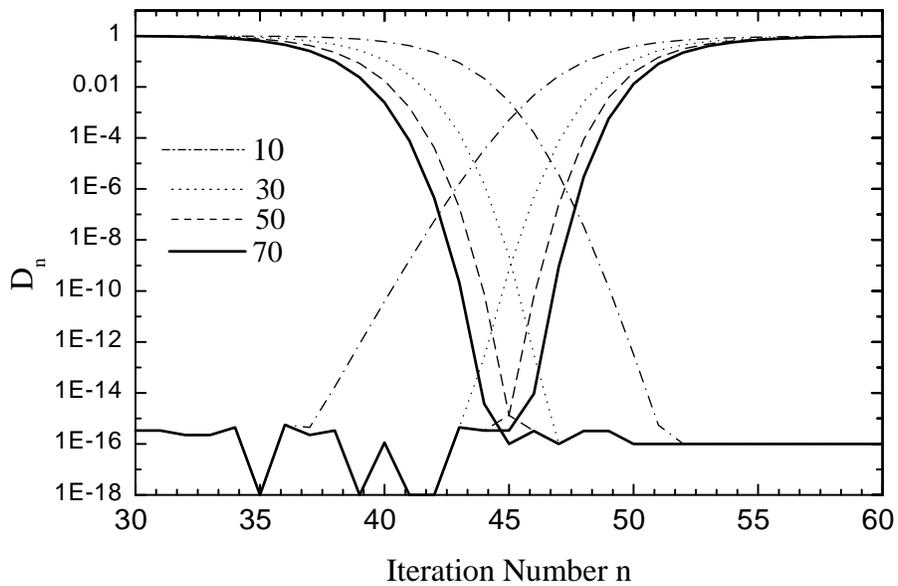,width=5in}}
\vskip125pt
\caption{Departure from scaling $D_n$ defined in the text, for $y$ (curves
reaching 1 to the right) and $\tilde{y}$ (curves reaching 1 to the left).
In each cases, we have considered approximations of order 10 (dot-dashes),
30 (dots), 50 (dashes) and 70 (solid line). The value of $\lambda$ is 1.5. }
\label{fig:tmoverlap}
\end{figure}

The relevant facts about the HM and its
RG transformation are reviewed in Section \ref{sec:hm}.
Before embarking in a multivariable calculation, we have 
first considered 
a simplified 
version involving only one variable\cite{dual} where the small denominator 
problem 
is obviously absent.
This model discussed in Section \ref{sec:tm}, 
is simply a quadratic map with two fixed points, one stable
and one unstable.
Series expansions for the scaling fields of this simplified model, 
and their inverse, can easily be 
constructed. Treating these numerical series with well-known 
estimators \cite{nickel80}, we obtain 
radii of convergence and exponents 
in very good agreement with what we can infer 
by using general arguments.
These numerical results will be used as references
when a similar analysis is conducted later for the HM.
The most important result of
Section \ref{sec:tm} 
is illustrated Fig. \ref{fig:tmoverlap} which shows that the 
expansions of the scaling
fields scale accurately in overlapping regions.

The rest of the article is devoted to the HM  with one spin component 
per site (as in the Ising model). 
For this model, the existence of a non-trivial unstable
fixed point has
been proven rigorously \cite{bleher75,collet78}.
Significant results have been obtained regarding the {\it local} existence
of scaling fields for 
the RG map near the unstable fixed point 
by Collet and Eckmann \cite{collet78} and Koch and Wittwer \cite{kw88}.
In addition, Ref. \cite{kw88} contains a 
mathematical justification of the polynomial approximations that we have used
to perform large order HT expansions \cite{osc1,osc2} or direct numerical
calculations \cite{finite,gam3}. 
In the following, the parameter playing the 
role of the dimensionality (see Section \ref{sec:hm}) 
will be tuned in such way that 
a Gaussian massless field scales exactly as in three dimensions.
In other words, we will work at an intermediate value between the upper
and lower critical dimensions and the $\epsilon$-expansion is not 
obviously useful.
For this particular choice of the dimensionality parameter, 
the numerical value of the unstable fixed point is known with great precision
in a specific system of coordinates \cite{koch95} and the question of 
small denominators studied in Ref. \cite{smalld}.

\begin{figure}
\centerline{\psfig{figure=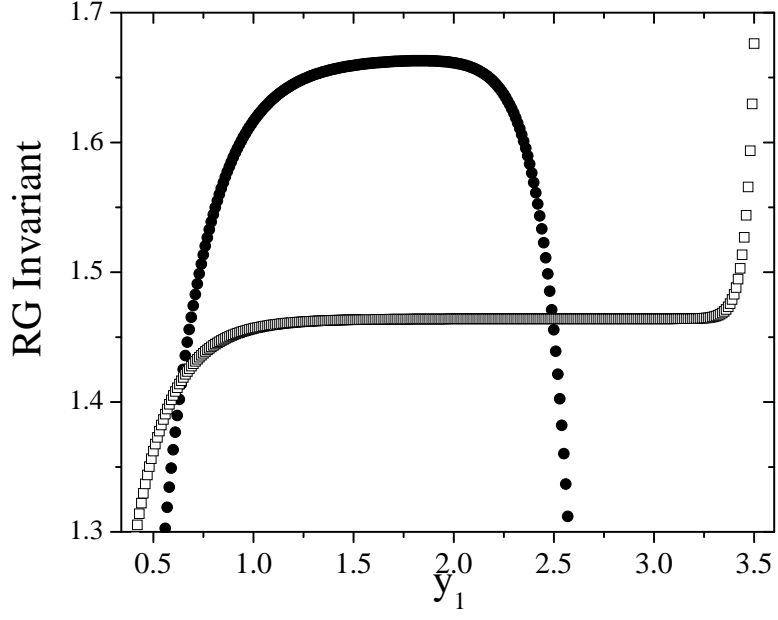,width=5.0in}}
\vskip125pt
\caption{Values of $C_1$ (empty squares) 
and $C_2$ (black circles) defined by Eq. (\ref{eq:rginv})}, as 
a function of $y_1$.
\label{fig:rginv}
\end{figure}

Using these results we first provide an explicit construction
of the two sets of scaling fields and show that the first two 
steps 
can be implemented. 
We found simple linear relations between the derivatives of the free 
energy and the HT scaling fields.
The RG invariants discussed above can be calculated accurately (see  
Fig. \ref{fig:rginv})
because the scaling fields 
associated with the two fixed points scale accurately
in overlapping regions as shown in Fig. \ref{fig:hmoverlap}. 
The third step was performed in the case of an initial Ising measure 
where the temperature is the only free parameter. Putting 
everything together, we can calculate the leading and subleading
amplitude of the susceptibility and have
very good agreement 
with previous numerical calculations \cite{gam3}.
This demonstrates the feasibility of the proposed startegy. 
In the conclusions, we discuss 
our plans to extend this method for perturbative initial
conditions and for nearest neighbor models.

A few words of caution. 
We want to make clear that the main result presented in this article
is a calculation of the leading and subleading  critical amplitudes of 
the magnetic susceptibility of the hierarchical model. This calculation 
shows that the strategy discussed in the coming section II 
can be implemented successfully for this particular model. 
The fact that part of the introduction and 
Section II are written for a general spin model should not
be interpreted as the statement that it is straightforward to implement
the strategy for other models than the hierarchical model. The reader whose
main interest is the hierarchical model may read section II as a general
description of what is done in the rest of the paper.
On the other hand, the reader interested in using the general strategy 
for other models, should be aware that we do not provide a general
procedure for constructing the unstable fixed point and its scaling 
fields. In addition, the fact that we found overlapping domain of 
convergence in the particular computation presented here is 
an encouraging result but it does not 
guarantee that a similar overlap will be found for other models.

\section{Three steps towards the calculation of the critical amplitudes}
\label{sec:3steps}

In this section, 
we consider a scalar model on 
a $D$-dimensional lattice with a lattice spacing $a_0$. 
We assume that $\beta < \beta_c$ and that the free energy density
$f$ is finite in the thermodynamic limit.  
We discuss the estimation of the critical 
amplitudes in the HT phase for $\chi^{(l)}$, the $l$-th derivative 
of $-f$ 
with respect to an external magnetic field. For definiteness, we generalize
the parametrization of Eq. (\ref{eq:param}) to
\begin{equation}
\chi^{(l)}= (\beta _c -\beta )^{-\gamma^{(l)} } (A_0^{(l)} + A_1^{(l)} 
(\beta _c -\beta)^{
\Delta^{(l)} }+\dots )\ .
\label{eq:paramgen}
\end{equation}
We assume the existence of a discrete RG transformation ${\bf R}$, 
which can be performed in the following way.
We first
integrate the fields in blocks of side $ba_0$ while keeping
the sum of the fields in the block constant.
We then rescale the sums of the fields by a factor 
$b^{(-2-D+\eta)/2}$. 
For the HM, $b=2^{1/D}$ and $\eta=0$.
We assume that this RG transformation has one 
non-trivial unstable fixed point and an attractive HT fixed point.
We first introduce general notations for the scaling fields and 
then discuss the three steps.

\subsection{Construction of the scaling fields}

A preliminary requisite is the construction of the 
two sets of scaling fields.
We construct the scaling fields near the unstable fixed point, denoted  
${\bf y}({\bf d})$, in terms 
of the coordinates ${\bf d}$ in 
the directions of the eigenvectors of the linearized RG map at that 
fixed point. 
The boldface notations mean that the quantity is a vector. 
In the following we will consider finite dimensional approximations
for such vectors.
For small values of $ {\bf d} $, we have 
${\bf{y}}({\bf d})\simeq {\bf d}$. One could call the ${\bf d}$,
the ``linear scaling fields'' or the ``real fields'' \cite{wegner72} 
if they have a simple 
physical interpretation. 
If we denote the RG transformation in the ${\bf d}$ coordinates as
${\bf  R}({\bf d})$, and ${\lambda_j}$ as the eigenvalue 
in the $j$-th direction, we have by definition 
of the scaling fields
\begin{equation}
y_j({\bf R}({\bf d}))=\lambda_j y_j({\bf d}).
\label{eq:scaw}
\end{equation}
In the following, we assume that the spectrum is real, positive,
non-degenerate and that only $\lambda_1>1$.

Similarly, the HT scaling fields denoted ${\bf \tilde{y}}({\bf h})$, 
can be constructed as expansions in 
the coordinates ${\bf h}$ in 
the directions of the eigenvectors of the linearized RG map at the 
HT fixed point. 
With notation analog to Eq. (\ref{eq:scaw}), we have
\begin{equation}
\tilde{y}_j({\bf R}({\bf h}))=\tilde{\lambda_j} \tilde{y}_j({\bf h}).
\label{eq:scat}
\end{equation}
The ${\bf d}$'s can be expressed in terms of the ${\bf h}$'s 
by a shift followed
by a linear transformation. In the following 
$y_{n,l}$ or $y_l({\bf d}_n)$ 
will denote the value of the $l$-th scaling field after $n$ 
iterations, and similar notations will be used for the HT variables.

\subsection{step 1}

The first step consists in expressing the $\chi^{(l)}$ in terms of 
the scaling fields of the HT fixed point ${\bf \tilde{y}}$.
In the vicinity of the HT fixed point, the 
effective lattice spacing becomes larger than
the physical correlation lengths. This fixed point is attractive 
and as the number of iterations $n$ becomes large, one can treat
the blocks as almost isolated systems with a large volume $b^{nD}$.
In the following we call $\chi^{(l)}_n$ the average value of the $l$-th 
power of the total spin in this volume minus its disconnected parts, divided 
by the volume.
Assuming that ${\rm lim}_{n\rightarrow\infty}\chi^{(l)}_n=\chi^{(l)}$ 
is finite for $\beta < \beta _c$, we conclude 
that the subtracted average value of the $l$-th power of the 
{\it rescaled} sum of all the spins scales like $b^{-((2+D-\eta)(l/2)-D)n}$
for $n$ large and should be expressible 
as products of the ${\bf \tilde{y}}$ with product of eigenvalues 
$b^{-((2+D-\eta)(l/2)-D)}$.
Indeed, in explicit calculations for the HM model, we found that 
simple linear relations hold, namely, when $n$ becomes large
\begin{equation}
\chi^{(2q)}_n\propto \tilde{y}_q({\bf h}_{n})(\tilde{\lambda_q})^{-n} \ ,
\end{equation}
and consequently,
\begin{equation}
\chi^{(2q)}=K^{(q)}\tilde{y}_q({\bf h}_{in})\ ,
\end{equation}
with ${\bf h}_{in}$ denoting the initial values and 
$K^{(q)}$ a constant depending on the choice the scales of the
coordinates ${\bf h}$ and easily calculable.
This is discussed explicitly in  subsection \ref{subsec:hmstep1} for the HM.
For the simplicity of the exposition, we will assume that this is 
also true for other models. Nevertheless it is straightforward 
to generalize the construction for the case of products of HT scaling 
fields.

\subsection{Step 2}

The expansion of $ \tilde{y}_q({\bf h}_{in})$
is expected to converge for ${\bf h}_{in}$ small enough, however
it might not be very useful or even meaningful near the unstable 
fixed point or the Gaussian fixed point.
If $\gamma^{(2q)} $denotes the leading critical exponent for $\chi^{(2q)}$, 
it follows from arguments based on the linear RG transformation that 
$\gamma^{(2q)}$ =\ -\ ln$\tilde{\lambda_q}$/ln$\lambda_1$.
Consequently,
$y_1^{\gamma^{(2q)}} \tilde{y}_q$
is RG-invariant. We can then factor out the susceptibility into a singular
part and a RG invariant part:
\begin{equation}
\chi^{(2q)}=K^{(q)}\lbrack y_1({\bf d}({\bf h}_{in})))\rbrack 
^{-\gamma^{(2q)}}\Biggl[ \lbrack
y_1({\bf d}({\bf h}_{in}))\rbrack^{\gamma^{(2q)}}
\tilde{y}_q({\bf h}_{in})\Biggr] \ .
\label{eq:gensus}
\end{equation}
The factor in large brackets is RG-invariant and does not need 
necessarily to be evaluated 
for initial values of ${\bf h}$. We can use the $n$-th iterate of these 
values ${\bf h}_n={\bf R}^n ({\bf h}_{in})$ for any $n$ and get the same answer. 
In particular, we can choose $n$ is such a way that the flow
is ``in between'' the two fixed points considered here, in a 
crossover region 
where the two expansions have a 
chance to be valid.
One of the main results presented in this article are 
numerical evidences that this procedure actually works. In other words,
that there exists an overlap between the domains where approximate expansions
of the scaling fields scale as they should. 

The scaling properties 
of expansions are related to convergence issues which need to 
be discussed in the complexification of the construction.
If an eigenvalue appearing in the defining equation for the scaling fields
Eqs. (\ref{eq:scaw}) and (\ref{eq:scat}), is such that 
$|\lambda_l|\neq
1$, it is clear that the only values that can be taken by $|y_l|$ 
at a fixed point
are 0 and $\infty$. Consequently, a detailed analysis of the fixed points
for the complexification of the RG transformation can put restrictions on
the domain of convergence of the scaling fields. Later, the inverse
functions ${\bf d}({\bf y})$ or ${\bf h}({\bf \tilde{y}})$ will also 
be used. Their radius of convergence can be restricted by the study of 
the extrema of the original function. In the case of 
the one dimensional model of Section \ref{sec:tm}, such a study can be 
conducted easily and confirms the results of the numerical analysis.

Near criticality, for the values of $n$ in the crossover discussed above, 
the values
of the scaling fields associated with the irrelevant directions are 
usually very small (unless very large initial values have been chosen)
and can be treated perturbatively. 
In first approximation, we can consistently set the initial values of
the irrelevant scaling fields to zero since they are multiplicatively
renormalized. In this approximation, we 
describe the flow along the unstable manifold.
A local calculation involving a scaling field corresponding to the 
the relevant direction is provided in Ref. \cite{kw88} following
a procedure developed in Ref. \cite{ew87} to prove
Feigenbaum conjectures.

It should also be noted that since the RG considered here is discrete 
(it is constructed by iterating ${\bf R}$ an integer number of times),
RG-invariant does not mean independent of ${\bf h}_{in}$. 
There is room for log-periodic 
corrections, which have been first noticed by Wilson \cite{wilson71b}, 
discussed in general
in Ref. \cite{niemeijer76} 
and observed in the HT expansion of the HM in Refs. \cite{osc1,osc2}.
These corrections are studied for the simplified model in Section 
\ref{sec:tm}. In our numerical study of the HM of Section \ref{sec:hm},
these corrections are 
too small to be resolved and will be ignored. 

\subsection{Step 3}

In the previous step, we traded a difficult problem (the estimation of 
the initial values of ${\bf \tilde{ y}}$) for two simpler problems: 
the estimation of the RG invariant and the estimation of  
the initial values of ${\bf y}$. Up to now, we have treated these
initial values as free variables
and constructed functions of these variables which depended only on
the RG transformation. We now need to incorporate the information
related to the actual initial values. This calculation depends on the 
type of models considered. For instance, for a Ising model, one expects 
\begin{equation}
y_1=Y_{1;1}(\beta_c-\beta)+Y_{1;2}(\beta_c-\beta)^2+\dots \ ,
\label{eq:init1}
\end{equation}
and 
\begin{equation}
y_l=Y_{l;0}+Y_{l;1}(\beta_c-\beta)+\dots \ ,
\end{equation}
for $l>1$. The $Y_{l;k}$ are constants that we will evaluate numerically for 
the HM. The non-leading terms are responsible for the analytical corrections
and are usually difficult to resolve numerically.
A more complete construction of 
the initial values in terms of the 
``bare parameters'' used 
in field theoretical perturbative calculations 
(with flows starting near the Gaussian fixed point) 
is in progress \cite{lilipro} but will not be discussed here.

\section{DYSON'S HIERARCHICAL MODEL}
\label{sec:hm}

In this section, we review the basic facts about the RG transformation
of the HM to be used in the rest of the paper and discuss various way 
to obtain finite dimensional truncations. For more detail, the reader
may consult Ref. \cite{collet78,finite} and other papers quoted in the 
introduction.
The energy density of the HM
has two parts. One part is non-local (the ``kinetic term'') and invariant 
under a RG transformation. Its explicit form can be found, for instance, 
in Ref. \cite{collet78} or 
in Section II of Ref. \cite{hyper}. The other part is a sum of 
local potentials given in terms of a unique function $V(\phi)$. The
exponential ${\rm e}^{-V(\phi)}$ 
will be called the local measure and denoted $W_0(\phi)$.
For instance, for Landau-Ginsburg models, the measures are of the
form $W_{0}(\phi)= e^{-A \phi^2-B\phi^4}$, but we can also consider limiting 
cases such as 
a Ising measure $W_{0}(\phi)=\delta (\phi^2-1)$.  
Under a block spin transformation which integrates 
the spin variables in  ``boxes'' with two 
sites, keeping their sum constant, the local measure transforms according to 
the integral formula
\begin{equation}
W_{n+1}(\phi) = \frac{C_{n+1}}{2} e^{(\beta/2)(c/4)^{n+1}\phi^2}
\int d \phi' W_{n} \left(\frac{ \phi - \phi'}{2} \right) W_{n} \left(
\frac{\phi + \phi'}{2} \right) \ , 
\end{equation}
where $\beta$ is the inverse temperature 
(or the coefficient in front of the kinetic term) 
and $C_{n+1}$ is a normalization
factor to be fixed at our convenience.

We use the Fourier transform
\begin{equation}
W_{n}(\phi) = \int \frac{d k}{2 \pi} e^{i k \phi} \hat{W}_{n}(k) \ ,
\end{equation}
and introduce a rescaling of $k$ by a factor $u/s^{n}$, where $u$ and $s$ are
constants to be fixed at our convenience, by defining
\begin{equation}
R_{n}(k) \equiv \hat{W}_{n}(\frac{u k}{s^{n}}) \ ,
\end{equation}
In the following, we will use $s=2/\sqrt{c}$ with $c=2^{1-2/D}$. This 
corresponds to the scaling of a massless gaussian field in $D$ dimensions.
Contrarily to what we have done in the past, we will here absorb the 
temperature in the measure by setting $u=\sqrt{\beta}$.
With these choices, the RG transformation reads
\begin{equation}
R_{n+1}(k) = C_{n+1} \exp \left[ -{1\over 2} 
{{\partial ^2} \over 
{\partial k ^2}} \right]\left[R_{n} \left({\sqrt{c}k\over 2} 
\right) \right]^2 \ . 
\label{eq:rec}\end{equation}
We fix the normalization constant $C_{n}$ so that $R_{n}(0) = 1$.
For an Ising measure, $R_{0}(k) = \cos(\sqrt{\beta}k)$, while in general,
we have to numerically integrate to determine the coefficients of
$R_{0}(k)$ expanded in terms of $k$.

If we Taylor expand about the origin,
\begin{equation}
R_{n}(k) =1+ \sum_{l=1}^{\infty} a_{n,l} k^{2 l} \ ,
\end{equation}
the finite-volume susceptibility reads
\begin{equation}
\chi_{n} = - 2 \frac{a_{n,1}}{\beta} \left(\frac{2}{c} \right)^{n} \ .
\label{eq:sus}
\end{equation}
The susceptibility $\chi$ is defined as
$\chi \equiv \lim_{n \rightarrow \infty} \chi_{n}$ .
For $\beta$ larger than
$\beta_{c}$, the definition of $\chi$ requires a subtraction 
(see e. g., Ref. \cite{hyper} for a practical implementation). 
In the following, we will only consider the 
HT phase ($\beta<\beta_c$).
The explicit form for $a_{n+1,l}$
in terms of $a_{n,l}$ reads
\begin{equation}
a_{n+1, l} = \frac{u_{n,l}}{u_{n,0}} \ ,
\label{eq:aofu}
\end{equation}
where
\begin{equation}
u_{n,l} \equiv \sum_{i=0}^{\infty} \frac{(- \frac{1}{2})^{i} 
(2(l+i))!}{s^{2(l+i)} i! (2 l)!}
 \sum_{p+q=l+i} a_{n, p} a_{n, q} \ .
\label{eq:recursion}\end{equation}

To study the susceptibility not too far from the HT
fixed point, we can expand $\chi$ in terms of $\beta$.
Since  we choose the scaling 
factor $u$ so that $\beta$ is eliminated from the recursion,
we find that $a_{0,l} \propto \beta^l$.  From the form of the
recursion, Eq.~(\ref{eq:recursion}), we can see that $a_{n,l}$
will always have $\beta^l$
as the leading power in its HT expansion (since $p+q\geq l$).  
We define
the coefficients of the expansion of the infinite-volume susceptibility by
\begin{equation}
\chi(\beta) = \sum_{m=0}^{\infty} b_{m} \beta^m \ .
\end{equation}
We define $r_{m} \equiv b_{m}/b_{m-1}$, the ratio of two successive
coefficients, and introduce quantities \cite{nickel80},
called the extrapolated
ratio ($\hat{R}_{m}$) and the extrapolated slope ($\hat{S}_{m}$)
which will be used later in a more general context and are  defined
by
\begin{equation}
\hat{R}_{m} \equiv  m r_{m} - (m-1) r_{m-1} \ , 
\label{eq:rhat}
\end{equation}
and
\begin{equation}
\hat{S}_{m} \equiv  m S_{m} - (m-1) S_{m-1} \ , 
\label{eq:shat}
\end{equation}
where
\begin{equation}
S_{m} \equiv  \frac{-m(m-1) (r_{m} - r_{m-1})}{m r_{m} - (m-1)r_{m-1}}   \ , 
\end{equation}
is called the normalized slope.
If we calculate $\hat{S}_{m}$ for the HM, 
we find oscillations illustrated in Refs.~\cite{osc1,osc2}.

The HT expansion can be calculated to very high order,
however, due to the amplification of some subleading corrections 
by the estimators,
this is an inefficient way to obtain information 
about the critical behavior of the HM. 
In Ref. \cite{finite}, it was found that one 
can obtain much better results by neglecting the contribution of 
the $a_{n,l}$ when $l\geq l_{max}$, with $l_{max}$ much smaller than the 
order of the HT expansion. As an example, one can calculate the 1000-th
HT coefficient of $\chi$ 
with 16 digits of accuracy using only 35 terms in the sum. 

This is equivalent to consider the polynomial
approximation
\begin{equation}
R_{n}(k) \simeq \sum_{l=0}^{l_{max}} a_{n,l} k^{2 l} \ ,
\end{equation}
for some integer $l_{max}$.
There remains to decide if one should or not 
truncate to order $k^{2l_{max}}$
after squaring $R_n$. This makes a difference 
since the exponential of
the second derivative has terms with arbitrarily high order
derivatives. Numerically, one gets better results at intermediate
values of $l_{max}$ by keeping all the terms in $R_n^2$.
In addition, for the calculations performed later, the intermediate truncation 
pads the ``structure constants'' of the maps (see sec. \ref{sec:multi}) with
about fifty percent of zeroes. 
A closer look at Section \ref{sec:multi}, may convince the reader that 
not truncating after 
squaring is more natural because we obtain 
correct (in the sense that they keep their 
value when $l_{max}$ is increased) structure constants in place of these 
zeroes. We have thus followed the second possibility where we truncate 
only once at the end of the calculation. With this choice
\begin{equation}
u_{n,l} \simeq \sum_{i=0}^{2l_{max}-l} \frac{(-\frac{1}{2})^{i} (2(l+i))!}
{(4/c)^{(l+i)} i! (2 l)!} \sum_{p+q=l+i} a_{n, p} a_{n, q} \ .
\label{eq:truncation2}\end{equation}

Compared to the HT expansion, the initial truncation to order $l_{max}$
is accurate up to order $\beta^{l_{max}}$. After one iteration, we 
will miss terms of order $\beta^{l_{max}+1}$ but we will also generate
some of the contributions of order $\beta^{2l_{max}}$ (but 
not all of them). After $n$ iterations
we generate some of the 
terms of order $\beta^{2^nl_{max}}$ as in superconvergent 
expansions. Rigorous justifications of the polynomial truncation 
can be found in Ref. \cite{kw88}.

\section{A ONE-VARIABLE MODEL}
\label{sec:tm}

Before attacking the multivariable expansions of the scaling fields, we
would like to illustrate the main ideas 
and study the convergence of series with a simple one variable example
which retains the important features: a critical
temperature, RG flows going from an unstable fixed point to
a stable one, and log-periodic 
oscillations in the susceptibility.

\begin{figure}
\centerline{\psfig{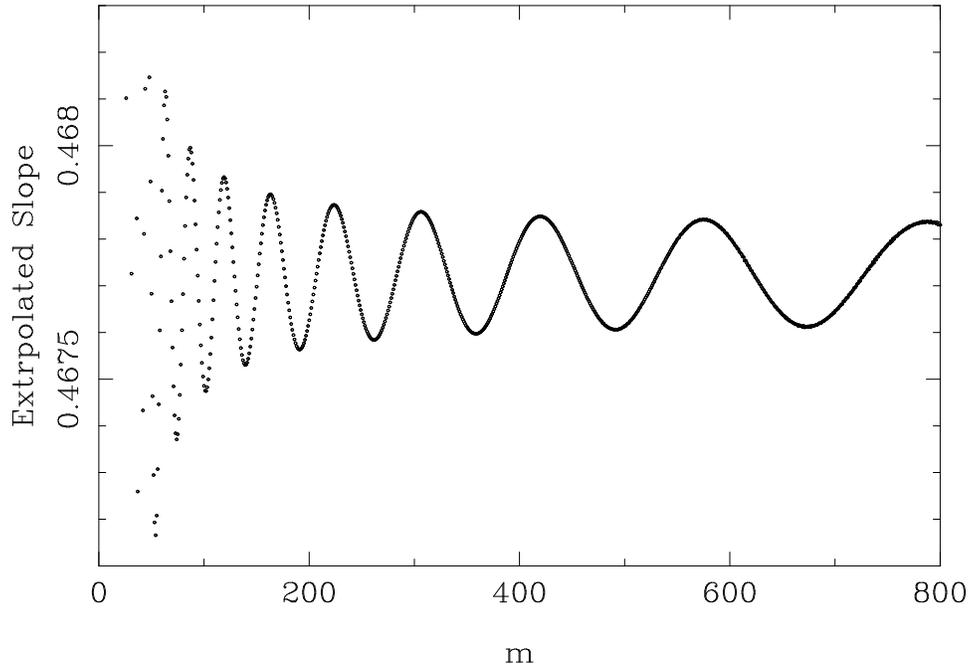}}
\vskip125pt
\caption{The extrapolated slope ($\hat{S}_m$) versus $m$ for the 
HT of $\chi$ calculated from the simplified recursion Eq. (\ref{eq:tmchi})
with $c=2^{1/3}$.}
\label{fig:tmexslope}
\end{figure}

In order to obtain a simple one-variable model, we first consider the
$l_{max} = 1$ truncation of
Eq. (\ref{eq:truncation2}):
\begin{equation}
a_{n+1,1} = \frac{ (c/2)a_{n,1} -(3c^2/8)a_{n,1}^2}
{1 - (c/2)a_{n,1} + (3c^2/16)  a_{n,1}^2} \ .
\end{equation}
Expanding the denominator
up to order $2$ in $a_{n+1,1}$
and using
Eq. (\ref{eq:sus}), we obtain
\begin{equation}
\chi_{n+1} = \chi_{n} + \frac{\beta}{4} \left(\frac{c}{2} \right)^{n+1}
\chi_{n}^{2} \ .
\label{eq:tmchi}
\end{equation}
This approximate equation was successfully  
used in Ref. \cite{finite} to model
the finite-size effects and was used as the 
starting point for a study of the scaling field in Ref. \cite{dual}.
If we expand $\chi$ (the limit of $\chi_n$ when $n$ becomes 
infinite) in $\beta$, and
define the extrapolated slope, $\hat{S}_m$, as in Eq.~(\ref{eq:shat}),
we see oscillations in Fig.~\ref{fig:tmexslope}
quite similar to those in the HM \cite{osc1,osc2}.
Using a rescaling discussed in Ref. \cite{dual} and the notation 
$\xi=c/2$, the map can be put in the 
canonical form
\begin{equation}
h_{n+1} = \xi h_{n} + (1-\xi) h_{n}^{2} \ .
\label{eq:hmap}\end{equation}
We call this map the ``$h$-map''. It has a stable fixed
point at 0 and an unstable fixed point at 1. We recover the susceptibility as:
\begin{equation}
\chi_{n} =  \frac{h_{n}}{h_{0}} \xi^{-n} \ 
\end{equation}

We can expand the map about the unstable fixed point, $h_{n} = 1$.
Using the new coordinate $d_{n} \equiv 1-h_{n}$ and the notation
$\lambda \equiv 2-\xi$, we obtain the ``dual'' map
\begin{equation}
d_{n+1} = \lambda d_{n} + (1-\lambda) d_{n}^{2} \ ,
\label{eq:dmap}\end{equation}
with the starting value $d_{0} = 1-\beta/\beta_{c}$.
We call this map the ``$d$-map'' and we can think of $d$ as being the 
distance to the critical point. We can construct   
a function $d$ such that $d(y_{n}) \equiv d_{n}$ by plugging an expansion
in $y_n$ into the equation
\begin{equation}
d(\lambda y_{n}) = \lambda d(y_{n}) + (1-\lambda) d^{2}(y_{n}) \ .
\end{equation}
Similarly, we can construct the inverse series 
and obtain $d(y_{n})$. This allows us 
construct $d_{n}$ in terms of $d_{0}$:
\begin{equation}
d_{n}=y^{-1}(\lambda^ny(d_0)) \ .
\label{eq:dsubn}
\end{equation}
Because of the duality between our two maps, we can
easily reproduce all of the above results of the $d$-map for
the $h$-map and express $h_{n}$ in
terms of a HT scaling field $\tilde{y}_{n}$.

We now turn to the three steps. Note that step 3 is not necessary: 
the knowledge of $y(d_0)$ provides the initial value of $y$ as a function of
$\beta$. Step 1 is straightforward.
As shown in Ref. \cite{dual}, in the infinite $n$ limit,
\begin{equation}
\chi = \frac{\tilde{y}_{0}}{h_{0}} = \frac{\tilde{y}(h_{0})}{h_{0}} \ .
\label{eq:suscep}\end{equation}
The only difficult part is step 2.
The susceptibility can be written as  
\begin{equation}
\chi = \frac{\Theta}{(1 - d_{0}) (y(d_{0}))^{\gamma}} \ .
\label{eq:chitm}
\end{equation}
with the RG invariant $\Theta\equiv\tilde{y}_{0} y_{0}^{\gamma}$.
Due to the discrete nature of our RG transformation, $\Theta$ 
is not exactly a constant. If expressed in terms of ln($y_0$),
$\Theta$ is a periodic function of period ln$\lambda$.
However, for $\lambda$ not too close to 2, the non-zero Fourier modes
are very small. Note also that the 
apparent singularity when $d_0\rightarrow 1$ is exactly canceled by 
$y(d_0)^\gamma $ by virtue of Eq. (\ref{eq:yya}) discussed below.
We now give empirical results concerning the large order
behavior of of the expansions of $y(d)$, $\tilde{y}(h)$ and their inverses. 

\begin{figure}
\centerline{\psfig{figure=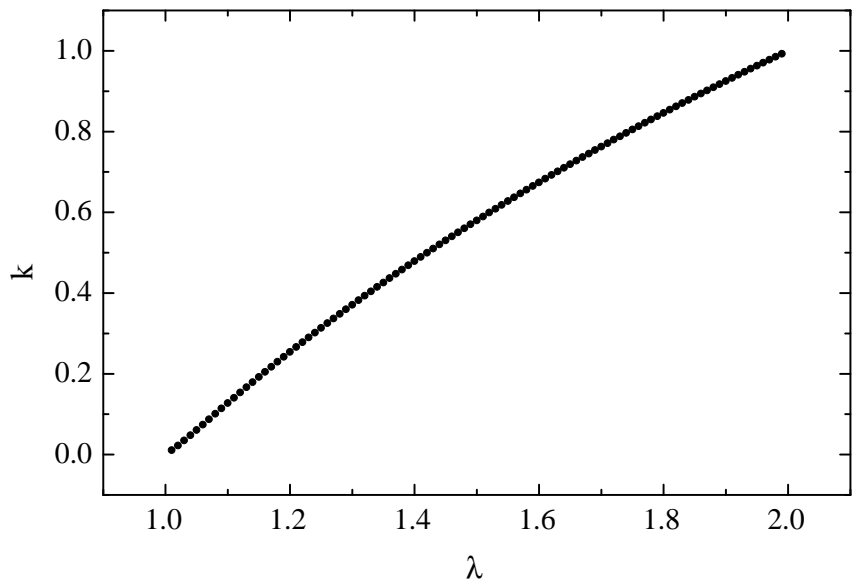,width=5in}}
\vskip125pt
\caption{The exponent $k$ defined in Eq. (\ref{eq:kexp}) as a function 
of $\lambda$. }
\label{fig:slope}
\end{figure}

We first consider $d(y)=y+\sum_{l=2}^{\infty}s_ly^l$.
For all tested values of $1<\lambda<2$, we obtained very good linear fits 
of $\ln |s_{l-1}/s_{l}|$ versus $\ln (l)$, for $l$ large
enough. Thus for large $l$, the coefficients obey the approximate rule
\begin{equation}
|\frac{s_{l}}{s_{l+1}}| \simeq C l^{k} \ ,
\label{eq:kexp}
\end{equation}
where we find that $C$ is always of order $1$ (in fact, $0.9<C<1$) and $0<k<1$.
Using iteratively this formula, we find 
that the coefficients decrease like $C^{-l}(l!)^{-k}$ and consequently
the $d(y)$ should be an entire function.
The numerical  values of $k$ are given in Fig. \ref{fig:slope}.

\begin{figure}
\centerline{\psfig{figure=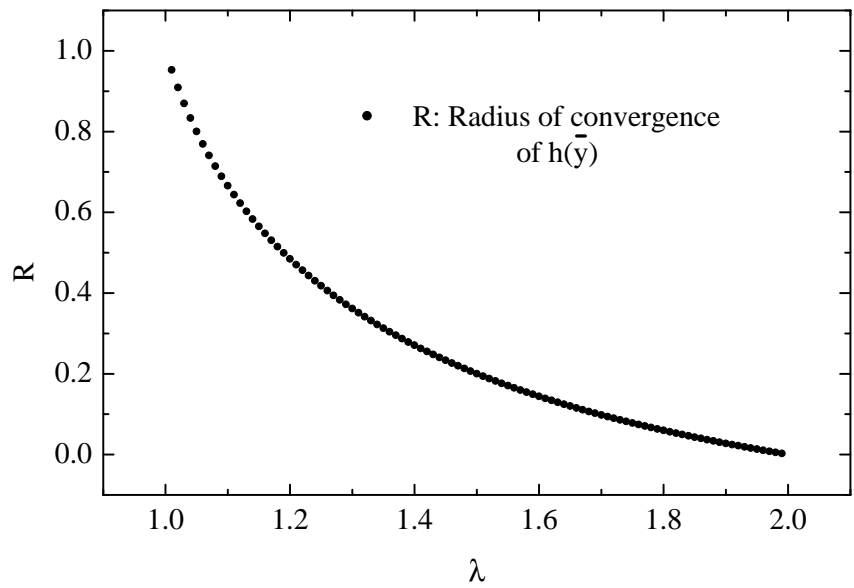,width=5in}}
\vskip125pt

\caption{Radius of convergence of $h(\tilde{y})$ as a function of 
$\lambda=2-\xi$. }
\label{fig:radius}
\end{figure}

We then consider $h(\tilde{y})$
and again examine the ratios of successive
coefficients, $|s_{l}/s_{l+1}|$.
We find the ratios flatten to constant values, for large enough
$l$,
indicating a finite radius
of convergence.  The radius get smaller
and vanish as $\xi$ approaches zero as shown in Fig. \ref{fig:radius}.
Note that for any value of $\xi$  tried, we found good evidence that
$lim_{n\rightarrow\infty}\widehat{S}_n=-1.5$, indicating a $(\tilde{y}-
\tilde{y}_c)^{1/2}$ behavior. This is consistent with the existence of
a quadratic minimum for the inverse function discussed below.

\begin{figure}
\centerline{\psfig{figure=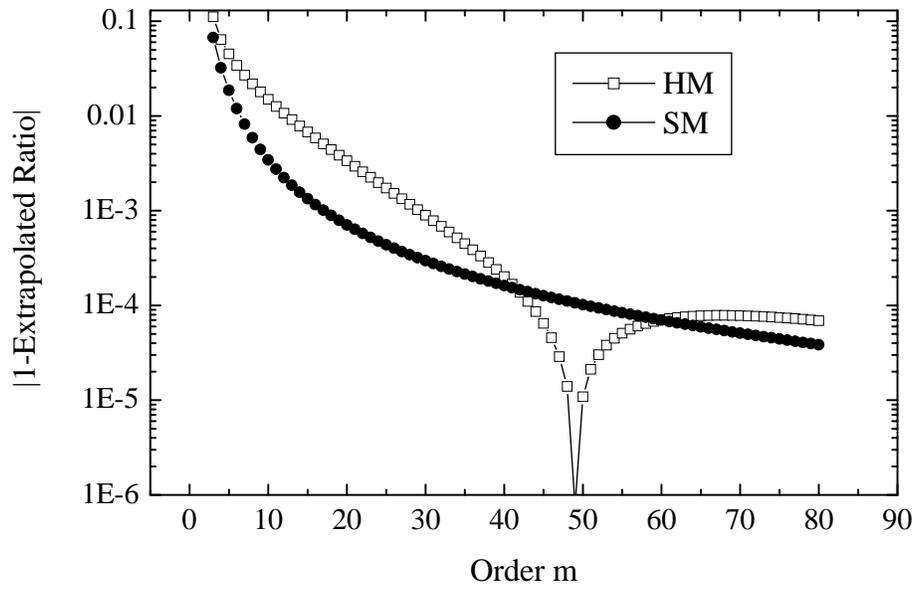,width=5.0in}}
\vskip125pt
\caption{The absolute value of the difference between the extrapolated ratio
and 1 for the HM (empty boxes) and the SM (full circles),
as a function of the order.}
\label{fig:rhdone}
\end{figure}

\begin{figure}
\centerline{\psfig{figure=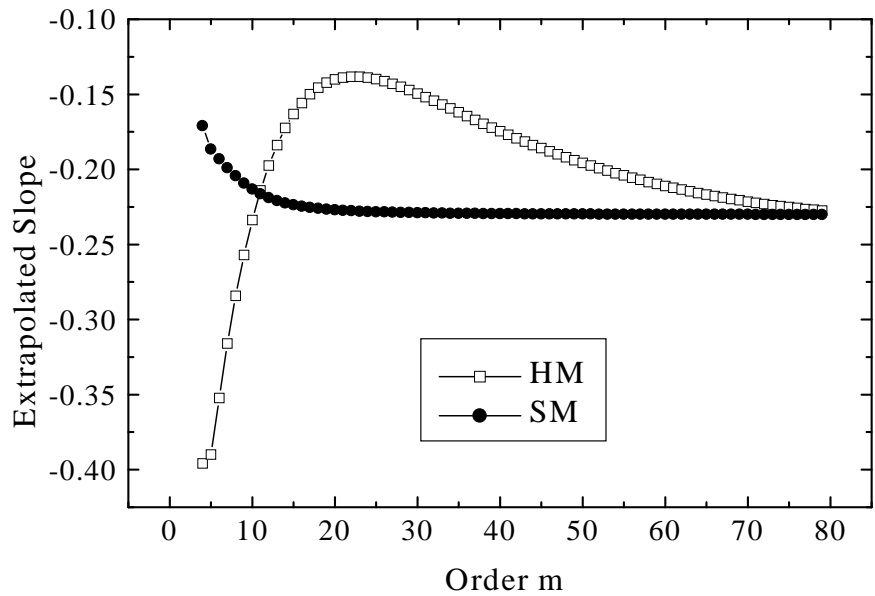,width=5.0in}}
\vskip125pt
\caption{The extrapolated slope $\hat{S}_m$ for the HM (empty boxes) 
and the SM (full circles) as a function of the order .}
\label{fig:shdone}
\end{figure}

We now turn to the inverse functions starting with $y(d)$.
As $d\rightarrow 1$, we reach the HT fixed point and we expect 
the convergence of the series to break down in this limit.
We find empirically from the analysis of ratios 
that for all $1<\lambda<2$, 
$y(d)$ converges in the region $0<d<1$.
The analysis of the extrapolated slope for various $\lambda$ gives 
convincing evidence that the main singularity has the form 
\begin{equation}
y(d)\sim (1-d)^{-1/\gamma}\ .
\label{eq:yya} 
\end{equation}
This can be seen with short series when 
$\lambda$ is close to one and requires larger and larger series as 
$\lambda$ gets close to 2.
Illustration of these properties for a particular value of $\lambda$ are
shown in Figs. \ref{fig:rhdone} and \ref{fig:shdone} where the estimators
are compared with those of the HM.

Finally, we discuss $\tilde{y}(h)$ which can be seen as 
a high-temperature expansion. 
We found clear evidence that 
the ratio of coefficients $t_{l}/t_{l+1}$ approaches $1$ for large $l$.
For smaller values of $\xi$, it takes larger order to 
reach this limit.
A detailed study shows that if we continue $\tilde{y}(h)$ for negative 
values of $h$ using the series expansion, the function develops a quadratic 
minimum at some negative value of $h$. The absolute value of $\tilde{y}$ 
at that value of $h$,
in all examples studied, reproduces accurately the radius of convergence
of the inverse function.
As one may suspect by looking at Fig. \ref{fig:tmexslope}, the 
analysis of the extrapolated slope 
is intricated. However, if we calculate enough terms and if $\lambda$ 
is not too close to 2, we get approximate results which
are consistent with a main singularity of the form 
\begin{equation}
\tilde{y}(h)\sim (1-h)^{-\gamma}\ .
\label{eq:yta}
\end{equation}
For instance, just by looking at 
the asymptotic behavior of 
Fig. \ref{fig:tmexslope}, one can see that $\gamma \simeq 1.4677$, 
as expected, with errors of the order $10^{-4}$.
It is interesting to note the duality \cite{dual} between Eqs. (\ref{eq:yya})
and (\ref{eq:yta}).

All the results concerning the convergence and the singularities have a 
simple interpretation. The 
finite radius of convergence of $y$ and $\tilde{y}$ is due to the other fixed
point which cannot be located inside the domain of convergence.
In this simple example there are exactly two fixed points in the 
complexification
of the map and this concludes the discussion. On the other hand,
the finite radius of $h$ is due to a minimum of $\tilde{y}$ at negative 
values  of $h$ while such a minimum does not appear for $d$.

As we have seen above, $\tilde{y}_{n} y_{n}^{\gamma} $ is 
independent
of $n$. We can thus pick $n$ such that we are just in the crossover
region and $both$ expansions have a reasonable chance to be accurate.
In order to test the accuracy of the approximations $y_{app}(d)$
(series expansion up to a certain order) for various $n$,
we have prepared an empirical sequences of $d_n$ starting with 
$d_0=10^{-8}$. We have then tested the scaling properties
by calculating
\begin{equation}
D_n=|\left[y_{app}(d_n)/(y_0\lambda^n)\right]-1|\ ,
\label{eq:bdn}
\end{equation}
where $y_0$ was calculated with 16 digits of accuracy
by using enough terms in the expansion of $y(d_0)$. For double precision 
calculations, 
optimal approximations are those for which $D_n\simeq 10^{-16}$.
For such approximation, the scaling is as good as it can possibly be
given the accuracy of $y_0$.
Indeed, due to the peculiar way numerical errors propagate \cite{numerr},
one does not reach exactly the expected level $10^{-16}$.
We can define a similar dual quantity by replacing $d$ by $h$ and
$y$ by $\tilde{y}$. In this case, $\tilde{y}_0$ is estimated with 
the same accuracy as $y_0$ by stabilizing $\tilde{y}(h_n)/\xi^n$,
for large enough $n$. 

We have performed  this calculation for $\lambda=$ 1.1, 1.5 and 1.9.
The conclusions
in the three cases are identical.
For $n$ large enough, the $D_n$ of $y$ starts increasing from $10^{-16}$
until it saturates around 1. 
By increasing the number of terms in the expansion, we can increase
the value of $n$ for which we start losing 
accuracy. Similarly, for $n$ low enough, the $D_n$ of $\tilde{y}$
starts increasing etc... We want to know if it is possible to 
calculate enough terms in each expansion to get scaling with some 
desired accuracy for both functions.
The answer to this question is affirmative 
according to Fig. \ref{fig:tmoverlap} for 
$\lambda =1.5$.
One sees, for instance, that with 10 terms in each series, we have 
scaling with about 1 part in 1000 near $n=45$ for both expansions.
The situation can be improved. For 70-70 expansions, an optimal accuracy
is reached from $n=44$ to 46. For the other values of $\lambda $ quoted
above, similar conclusions are reached, the only difference being
the optimal values of $n$.

\begin{figure}
\centerline{\psfig{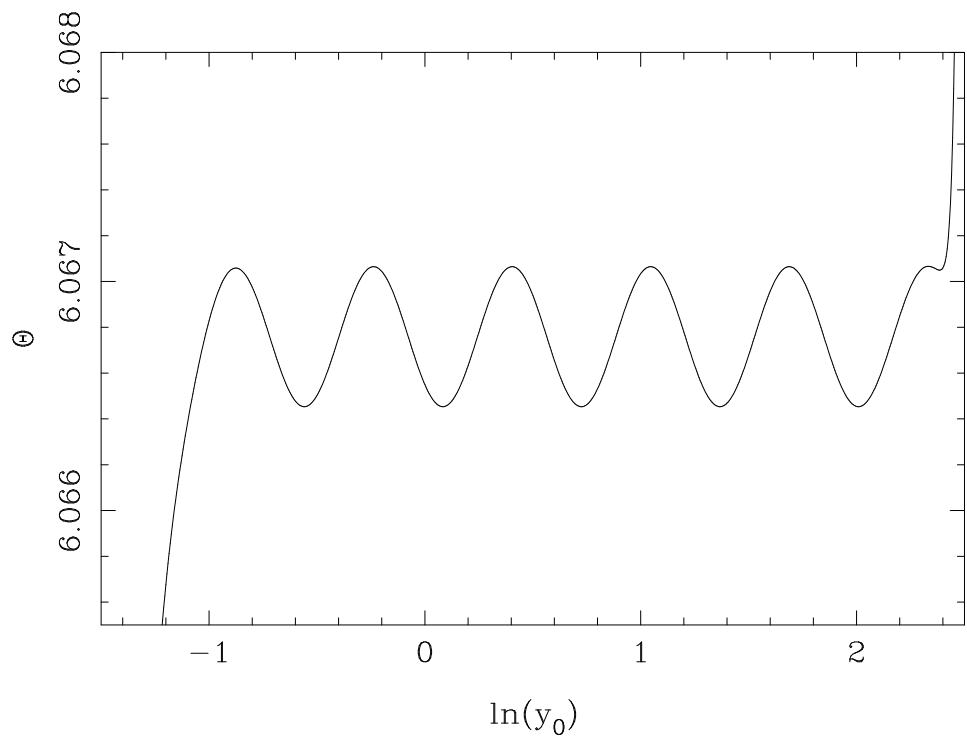}}
\vskip125pt
\caption{The invariant function $\Theta$, calculated at $\xi = 0.1$,
and plotted against $\ln(y_{0})$.}
\label{fig:all}
\end{figure}

Another evidence for overlapping convergence is that we can stabilize 
the RG invariant $\Theta$ for a certain range of $y_n$.
To evaluate $\Theta$, we use the series expansions for
$\tilde{y}$ and $d$, cutting each off at some order:
\begin{equation}
\tilde{y}(1 - d(y)) \simeq \sum_{i=1}^{\tilde{m}}
t_{i} (1 - \sum_{j=1}^{m} s_{j} y^j)^i \ ,
\end{equation}
where $s_{l}$ and $t_{l}$ are the $l$th coefficients in the $d$ and
$\tilde{y}$ series, respectively.
We have found that, given a fixed value of $m + \tilde{m}$, the most
accurate values for $\Theta$ are obtained when $m \simeq \tilde{m}$.
In Fig.~\ref{fig:all}, we show
$\Theta$ calculated by keeping $50$ terms
each in the expansions for $y$ and $\tilde{y}$.  The result is plotted
against $\ln(y)$. We used $\xi = 0.1$, which makes the oscillations
much larger than, for example, near to $\xi = 2^{-2/3}$.
Near the fixed points, we need more terms in the appropriate series
to get accurate results.

We can study the oscillation we see in $\Theta$ by looking at
its Fourier expansion.
Since $\Theta$ is periodic in $\ln y_{0}$, we can express
\begin{equation}
\Theta (y_0) = y_0^{\gamma}
\tilde{y}(1-d(y_{0})) = \sum_{p} a_{p} e^{i p \omega \ln y_{0}} \ ,
\end{equation}
where $\omega \equiv 2 \pi/\ln \lambda$.
The coefficients are given as
\begin{equation}
a_{p} = \frac{1}{\ln \lambda} \int_{y_{a}}^{\lambda y_{a}}
y^{\gamma-1-ip \omega} \tilde{y}(1 - d(y)) dy \ ,
\label{eq:fourier}\end{equation}
As an example, we calculated $a_{0}$ for $\xi = 0.1$ where the oscillations 
are not too small. The choice of the interval of integration 
can be inferred from 
Fig. \ref{fig:all}. 
If we had infinite series, the function would be exactly periodic.
For finite series, we see that $y_a$ cannot be too large or too small.
For intermediate values, we 
obtain $a_{0} \simeq 6.06676$. Proceeding similarly, we were able 
to resolve the next two Fourier modes. For reference, the magnitude 
of $a_2$ is about 
$2\times 10^{-10}$. Using $\Theta\simeq a_0$ together with 
Eq. (\ref{eq:chitm}), we obtain the leading amplitude together with 
the analytical corrections coming from the nonlinear tems in $y(d)$.

\section{SCALING FIELDS IN THE HIERARCHICAL MODEL}
\label{sec:multi}
\subsection{Construction of the scaling fields}

For notational convenience, we 
first rewrite the unnormalized recursion given in Eq. (\ref{eq:truncation2}),
using the ``structure constants'':
\begin{equation}
u_{n,\sigma} = \Gamma_{\sigma}^{ \mu \nu} a_{n,\mu} a_{n,\nu} \ ,
\end{equation}
with
\begin{equation}
\Gamma_{\sigma}^{ \mu \nu}
= \left\{ 
\begin{array}
{r@{\quad,\quad}l}
(c/4)^{\mu+\nu}\ 
\frac{(-1/2)^{\mu + \nu - \sigma}(2(\mu+\nu))!}{ 
(\mu+\nu-\sigma)!(2 \sigma)!}  \ &\   {\rm for}\  \mu+\nu \geq\sigma \\
0\ &\ {\rm otherwise}\ .
\end{array} \right.
\label{eq:struct}
\end{equation}
These zeroes can be understood as a ``selection rule'' 
associated with the fact that $a_{n,l}$ is of order $\beta^l$ 
as explained in Section \ref{sec:hm}.
If we follow the truncation procedure explained in Section \ref{sec:hm},
the indices simply run over a finite number of values.
We use ``relativistic'' notations. 
Repeated indices mean summation.
The greek indices indices $\mu$ and $\nu$
go from $0$ to $l_{max}$, while latin indices $i$, $j$ go from 1 to   
 $l_{max}$. Obviously,
$\Gamma_{\sigma}^{\mu \nu}$ is symmetric in
$\mu$ and $\nu$.
By construction, $a_{n,0} = 1$ and we can write the normalized 
recursion in the form:
\begin{equation}
a_{n+1,l} = \frac{{\cal M}_l^i a_{n,i}
 + \Gamma_{l}^{ i j} a_{n,i} a_{n,j}}
{1 + {\cal M}_0^i a_{n,i} + \Gamma_{0}^{i j} 
a_{n,i} a_{n,j}} \ ,
\label{eq:general}
\end{equation}
with
 ${\cal M}_{\eta}^{ i} = 2 \Gamma_{\eta}^{ 0 i}$ .

We then expand the basic map of Eq. (\ref{eq:general}) about the two 
fixed points of interest, choosing coordinates such that the 
matrix associated with the linearized RG
transformation becomes diagonal. This matrix is not symmetric and the 
relations of orthogonality and completeness 
are left invariant under rescaling of any right eigenvector 
by a nonzero constant together with a rescaling of the corresponding 
left eigenvector by the inverse. In the following, we will 
fix this ambiguity by requiring, in analogy with Section \ref{sec:tm} that 
the ``other'' fixed point (i.e. the one not located at the origin
by construction) be
at $(1,1,\ldots)$ in the new coordinates. 
 
The HT fixed point is at the origin of the coordinates in Eq. 
(\ref{eq:general}) and all we need to do is diagonalizing
the linear form  ${\cal M}_i^j$.
This is quite simple because it is of the upper triangular form.
The eigenvalues
are just the diagonal terms.  From Eq. (\ref{eq:struct}), 
one sees that $l$th diagonal term is
given as 
\begin{equation}
\tilde{\lambda}_{r}=2(c/4)^{r} \ .
\label{eq:hteigenv}
\end{equation}
This spectrum was obtained in Lemma 3.3 of  
Ref. \cite{collet78} with a different method.
This spectrum has a simple interpretation which will be discussed 
below. 
The upper diagonal form implies that the $l-th$ right eigenvector has 
only its $l$ first entries non-zero. This means that 
if we truncate to $a_{l_{max}}$, we are simply truncating
to a subspace spanned by the first $l_{max}$ eigenvectors. 
This is an interesting reinterpretation of the original polynomial
truncation which can be applied for other models.
Introducing new coordinates $h_{l}$, so that 
$a_{n,l} = \tilde{\psi}^r_l h_{n,r}$ with $\tilde{\psi}$ the matrix 
of right eigenvectors, the RG transformation has the form
\begin{equation}
h_{n+1,r} = \frac{\tilde{\lambda}_{r} h_{n,r}
 + \tilde{\Delta}_{r}^{ p q} h_{n,p} h_{n,q} }
{1 + \tilde{\Lambda}^{p} h_{n,p} 
+ \tilde{\Delta}_{0}^{ p q} h_{n,p} h_{n,q}}\ .
\end{equation}
Note that the form of the eigenvectors guarantees that 
$h_{n,l}$ is of order $\beta^l$. This can be seen by inverting 
the linear change of variable using the matrix of left eigenvectors.
Due to the upper-diagonal form of the linear transformation, 
the second left eigenvector has 
its first entry equal to zero, the third its first two entries etc... .

Near the nontrivial fixed point, we first use accurate values of 
the fixed point \cite{koch95} to bring the fixed point at the origin.
The eigenvectors are then calculated numerically using truncated 
forms of the linearized RG transformation. There is no exact 
closure as in the HT case, however the first 
eigenvalues stabilize rapidly when $l_{max}$ increases.
There is only one eigenvalue larger than $1$.
For instance the 
numerical values for $c=2^{1/3}$ are $\lambda_1=1.42717\dots$, 
$\lambda_2=0.85941\dots$. A more complete list 
is given in Ref. \cite{gam3}.
In summary, we can choose a system \
of coordinate $d_l$ where 
the unstable fixed point will be at the origin of the coordinate
and the 
HT fixed point at $(1,1,\dots)$, and such that the RG transformation has the 
form
\begin{equation}
d_{n+1,r} = \frac{\lambda_{r} d_{n,r}
 + \Delta_{r}^{ p q} d_{n,p} d_{n,q} }
{1 + \Lambda^{p} d_{n,p} + \Delta_{0}^{ p q} d_{n,p} d_{n,q} } \ ,
\label{eq:hmdrules}
\end{equation}

We can express 
canonical coordinates (linear scaling fields) in terms of 
the nonlinear scaling fields:
\begin{equation}
d_{n,r} = \sum_{i_{1},i_{2},\ldots} s_{r,i_{1} i_{2} \ldots} y_{n,1}^{i_{1}}
y_{n,2}^{i_{2}} \ldots \ ,
\end{equation}
where the sums over the $i$'s run from $0$ to infinity in each variable
and $y_{n,l}=\lambda_{l}^n y_{0,l}$.
Using the notation ${\mathbf{i}} = (i_{1},i_{2} \ldots$)
and the product symbol, we may rewrite the expansion as
\begin{equation}
d_{n,r} = \sum_{\mathbf{i}} s_{r,\mathbf{i}} \prod_{m} y_{n,m}^{i_{m}}
\end{equation}
Using the transformation law for the scaling fields, we have
\begin{equation}
d_{n+1,r} = \sum_{\mathbf{i}} s_{r,\mathbf{i}}  \prod_{m} (\lambda_{m}
 y_{m})^{i_{m}} \ .
\end{equation}
Each constant term, $s_{r,0,0,\ldots}$, is zero, as the scaling
fields vanish at the fixed point.  From Eq.~(\ref{eq:hmdrules}),
we see that all but one of the linear terms are zero for each value of $r$.  
The remaining term is the one proportional to the $r$th scaling variable.
We take these coefficients to be $1$, so that the $d_{n,r} \simeq y_{n,r}$
for small $y_{n,r}$.  For the higher-order terms, we obtain the recursion
\begin{equation}
s_{r, \mathbf{i}} = \frac{ \sum_{\mathbf{j}+\mathbf{k} = \mathbf{i}}
 ( \Delta_{r}^{ p q} s_{p,\mathbf{j}}
 s_{q,\mathbf{k}} - s_{r,\mathbf{j}} \prod_{m} \lambda_{m}^{j_{m}}
\Lambda^{p} s_{p, \mathbf{k}} ) -\sum_{\mathbf{j}
+\mathbf{k}+\mathbf{l}=\mathbf{i}}
 s_{r, \mathbf{j}} \prod_{m} \lambda_{m}^{j_{m}} \Delta_{0}^{ p q}
 s_{p, \mathbf{k}} s_{q, \mathbf{l}} }
{\left(\prod_{m} \lambda_{m}^{i_{m}} - \lambda_{r}\right)} \ .
\label{eq:sca}
\end{equation}
The calculation can be organized in such way that the r. h. s. of the 
equations are already known. This will be the case for instance if 
we proceed order by order in $\sum_q i_q$, the degree of non-linearity.
This expansion may have small
denominator problems. However, as discussed in the introduction, 
using numerical values of the eigenvalues
as calculated in \cite{gam3}, we did not find spectacular cancellations
between the two terms entering the denominator.

We can likewise expand each $h_{n,r}$ in terms of scaling fields
$\tilde{y}_{n,1},\tilde{y}_{n,2}, \ldots$.  The derived recursions
are identical in form to those derived above. From Eq. (\ref{eq:hteigenv}),
one sees that the denominator will vanish for some equations. This 
question is discussed in \cite{smalld} where it is shown that to all
order relevant for the following calculation, a zero denominator 
always comes with a zero numerator.

One can likewise find expansions of the scaling fields in terms of 
the canonical coordinates, by setting 
\begin{equation}
y_{n,r} = \sum_{\mathbf{i}} u_{r,\mathbf{i}} \prod_{m} d_{n,m}^{i_{m}}\ ,
\end{equation}
and requiring that when ${\bf d}_n$ is replaced by ${\bf d}_{n+1}$, the 
function is multiplied by $\lambda _{r}$. Since ${\bf d}_{n+1}$ has a
denominator, it needs to be expanded for instance in increasing order
of non-linearity. A simple reasoning shows that 
exactly the same small denominators as in Eq. (\ref{eq:sca}) 
will be present in these calculations. 
The same considerations applies for $\tilde{y}$.

\subsection{Practical implementation}
\label{subsec:prac}
We have calculated an empirical series of $a_{n,l}$ with
$c=2^{1/3}$, an initial Ising measure and 
$\beta=\beta_c-10^{-8}$ . 
Detail relevant 
for this calculations can be found in Refs. \cite{finite,gam3,numerr}.
The calculations have been performed with $l_{max}=30$, a value for
which at the $\beta$ considered, the errors due to the 
truncation are of the same order as those due to the numerical errors.
These errors are small enough to 
allow a determination of the susceptibility with seven
significant digits if we use double precision.

We now discuss the flow chronologically. 
Our choice of $\beta$ (close to $\beta_c$) 
means that we start near the stable manifold.
After about 25 iterations, we start approaching the unstable fixed point
and the linear behavior $d_{n+1,l}\simeq\lambda_l d_n$ becomes a good
approximation. During the next 20 iterations, the irrelevant variables
die off at the linear rate and at the same time the flow moves away from the 
fixed point along the unstable direction, 
also at the linear rate. At $n=47$ we are in good approximation on the 
unstable manifold and $d_{n,2}$ becomes proportional to $d_{n,1}^2$.
In other words, the non-linear terms are taking over.
At this point, we can approximate the $d_{n,l}$ as functions
of $y_1$ {\it only}: $d_{n,l}\simeq d_l(\lambda_1^n y_1,0,0,\dots)$. This 
approximation is consistent in the sense that if $y_2=0$ at $n=0$
then it is also the case for all positive $n$. 

We have calculated $d_l(y_1,0,0,\dots)$ up to order 80 in $y_1$ 
using Eq. (\ref{eq:sca}). 
There cannot be any small denominators in this restricted case.
We have then inverted $d_1$, 
now a function of $y_1$ only,  to obtain $y_1(d_1)$.
Given the empirical $d_{n,1}$ we then calculated the approximate
$y_{n,1}$ and then used the other functions $d_l(y_1)$ 
(with $l\geq2$) calculated before to ``predict''
$d_{n,l}$. Comparison with the actual numbers were good in a restricted 
range. For $n=49$, the relative errors were
less than a percent. They kept decreasing to less than one part in 
10,000 for 
$n=54$ and then increased again. It will be shown later 
that this corresponds to the fact that
when $y_1$ becomes too large (a value of approximately 
3.7 first exceeded at
$n=57$), the series expansion of $d_1$ seem to diverge, unlike the 
one-variable model for which $d(y)$ is an entire function.

The quality of the approximation between 
$n=45$ and $n=55$ can be improved  
by treating $y_2$ perturbatively. 
We have expanded 
\begin{equation}
d_l(y_1,y_2,0,\dots)
\simeq d_l^{(0)}(y_1,0,0,\dots)+y_2d_l^{(1)}(y_1,0,0,\dots)\ ,
\label{eq:pertexp}
\end{equation}
with $d_l^{(1)}$ up to order 30
in $y_1$. 
This allows us to obtain the first order expression:
\begin{equation}
\tilde{y}_1({\bf h}({\bf d}(y_1,y_2,0,\dots)))\simeq G(y_1)+y_2 H(y_1)\ .
\label{eq:paramy}
\end{equation}
Note that expansions at {\it both} fixed points 
are involved (one for $\tilde{y}_1$ and one for 
${\bf d}$) in this equation. When 
finite series are used, the approximation will only be valid in 
the crossover region.
Near $n=55$, the presence of the HT fixed point starts dominating
the flow but we are still far away from the linear regime. 
We have taken these non-linear effects in $\tilde{y}_1({\bf h})$
into account by calculating it 
up to order 11 in $\beta$. This is 
a multi-variable expansion. Recalling the discussion about the HT
expansion in 
Section \ref{sec:hm} and the properties of the 
eigenvectors of the linearized RG transformation about the HT fixed point, 
we can count the number of terms at each
order in $\beta$. At order two, we have $h_1^2$ and $h_2$, but since 
the linear transformation is diagonalized, $h_2$ will only appear in 
$\tilde{y}_2({\bf h})$ with coefficient 1. It is 
easy to see that at order $l$, one needs to determine $p(l)-1$ 
coeffificients, where
$p(l)$ is the number of partitions of $l$. It has been known from the 
work of Hardy and Ramanujan that
\begin{equation}
p(l)\sim \frac{{\rm exp}(\pi \sqrt{2l/3})}{4\sqrt{3} l}\ .
\end{equation}
It seems thus difficult to get very high order in this expansion.
In order to fix the ideas, there are 41 terms at order 11, 
489 at order 20 and 13,848,649 at order 80.

As we will explain below,
the expansion up to order 11 has a sufficient accuracy to be used 
starting 
at $n=55$. It also provides
optimal (given our use of double precision) results for $n\geq 60$.
As $n$ increases beyond 60, one can see the effect of each order
disappear one after the other as discussed in \ref{subsec:hmsec2}. 
Finally, the linear behavior becomes
optimal near $n=130$. 
This concludes our chronological discussion. 

\subsection{Step 1}
\label{subsec:hmstep1}
From Eqs. (\ref{eq:sus}) and remembering that we have absorbed 
$\beta$ in the $a_{n,}$, we obtain
\begin{equation}
\chi_{n} = -(2/\beta)\tilde{\psi}^r_1 h_{n,r} (2/c)^n \ ,
\end{equation}
where $\tilde{\psi}$ is the matrix of right eigenvectors.
For $n$ large enough, the linear behavior applies and the 
$h_{n,r}$ get multiplied by $2(c/4)^r$ at each iteration.
In the large $n$ limit, only the $r=1$ term survives and 
consequently,
\begin{equation}
\chi=  -(2/\beta)\tilde{\psi}^1_1 {\rm lim}_{n\rightarrow\infty}
h_{n,1} (2/c)^n \ .
\end{equation}
Using the same method as in the one-variable model, we 
can in the limit replace $h_{n,1}$ by $\tilde{y}_{n,1}$ and obtain
\begin{equation} 
\chi=  -(2/\beta)\tilde{\psi}^1_1 \tilde{y}_{0,1}\  .
\end{equation} 
For reference, in the case $c=2^{1/3}$ and with the normalization
of the eigenvectors discussed above, $\tilde{\psi}^1_1
\simeq -0.564$.
Also note that since for all $l$, 
$0<\tilde{\lambda}_l<1$, all other monomials in the 
$\tilde{y}_l$ go to zero faster than $\tilde{y}_1$.

One can calculate the subtracted 2$q$-point function following the same 
procedure. As shown in \cite{finite,hyper}, they can be expressed in terms
of $a_{q,n}$ and higher powers of the $a_{n,l}$ with $l<q$.
Following the procedure described above, these quantities can then 
reexpressed in terms of $\tilde{\bf y}$. We need to 
identify the leading term in this expansion.
By rescaling $(-1)^q (2q)!a_{q,n}$ by $(4/c)^n$ we obtain the average 
value of the $2q$-th power of the main spin (sum of the $2^n$ 
spin variables $\phi_x$). In the symmetric phase, 
this quantity scales like $2^{qn}$ when $n$ increases. However, the 
subtracted version of this quantity (which is generated by the free
energy) is expected to scale 
like $2^n$. In other words, if we assume that the free energy density 
is finite
\begin{equation}
a_{q,n}-({\rm subtractions})\propto \lbrack 2(c/4)^q\rbrack ^n \ .
\end{equation}
One clearly recognizes the spectrum of Eq. (\ref{eq:hteigenv}) and
expects that the leading term is 
\begin{equation}
a_{q,n}-({\rm subtractions})\propto \tilde{y}_{n,q}\ .
\end{equation}
In order to prove this conjecture by direct algebraic methods, 
one needs to show that the 
the non-linear terms 
which scale faster than $\tilde{y}_{n,q}$ are canceled by the subtraction.
For instance for the subtracted four-point function and  $c=2^{1/3}$,
$\tilde{\lambda}_1^2>\tilde{\lambda}_2$  
and $\tilde{\lambda}_1^3>\tilde{\lambda}_2$. 
We have checked 
that the corresponding terms $(\tilde{y}_{1})^2$ and $(\tilde{y}_{1})^3$
disappear with the subtraction. 
We have conducted similar checks for the 
6 and 8 point functions\cite{lilipro} and found similar 
cancellations.
It should be noted that a rigorous proof that the mechanism 
works in general, would imply the finiteness of the free energy 
density and hyperscaling (as defined in \cite{hyper}).
The practical calculation of the subtracted quantities is made difficult 
by the fact that as $n$ increases, the ``signal'' becomes much smaller 
than the ``background'' (the unsubtracted part). This requires the 
use of adjustable precision arithmetic. In the following, we will
only discuss the 2 point function (susceptibility).

\subsection{Step 2}
\label{subsec:hmsec2}
We rewrite the susceptibility as 
\begin{equation}
\chi\simeq (1.127853/\beta)\Theta(y_{0,1})^{-\gamma}\  ,
\label{eq:suspar}
\end{equation}
with the RG invariant $\Theta \equiv \tilde{y}_{0,1} y_{0,1}^{\gamma} =
 \tilde{y}_{n,1} y_{n,1}^{\gamma}$.
We first constructed $y_1(d_1)$ by neglecting the effects of the irrelevant
directions, as explained 
in subsection \ref{subsec:prac}. 
In order to provide a comparison, we have calculated the same
$y(d)$ for the one-variable model with $\lambda=1.2573$. 
In the following, we call this model the ``simplified model'' (SM).
For this
special value of $\lambda$, the critical exponents $\gamma$ of the two models
coincide with five significant digits. 
We have good evidence that both series have a radius 
of convergence 1 as indicated by the extrapolated ratio 
defined in Eq. (\ref{eq:rhat}) reaching
1 at an expected rate (Fig. \ref{fig:rhdone}).
Similarly, their extrapolated slopes seem to converge to the same value 
$1/\gamma -1\simeq -0.23026\dots $ as shown in Fig. \ref{fig:shdone}.
In conclusion, the function $y(d)$ for the SM is a 
reasonably good model to 
guess the asymptotic behavior of $y_1(d_1)$. 
Remembering the explanations of Section \ref{sec:tm}, this indicates the 
absence of other fixed points in the vicinity of the unstable fixed point.

\begin{figure}
\centerline{\psfig{figure=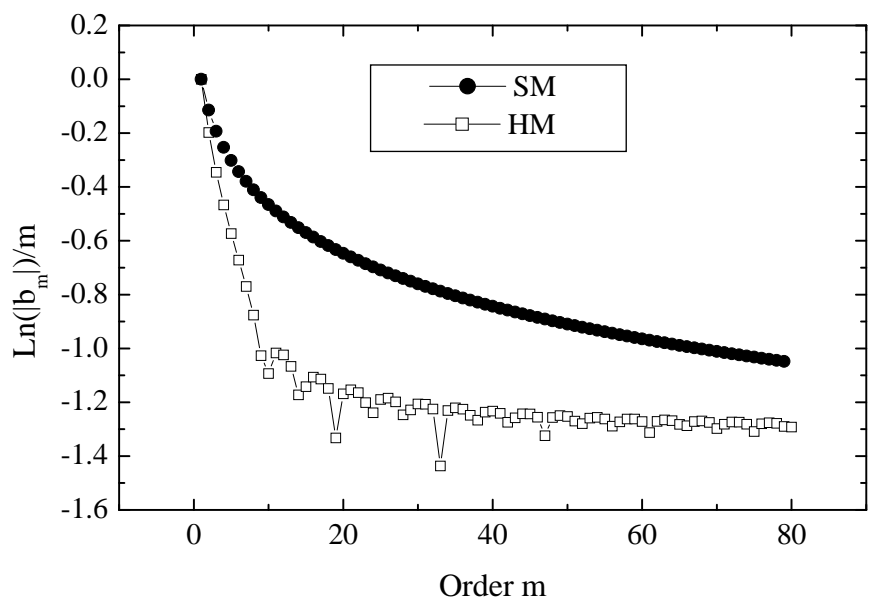,width=5.0in}}
\vskip125pt
\caption{Logarithm of the absolute value of the coefficients of the 
expansion of $d_1(y_1)$ divided by the order, for the HM (empty boxes) and the 
SM (circles).}
\label{fig:logco}
\end{figure}

\begin{figure}
\centerline{\psfig{figure=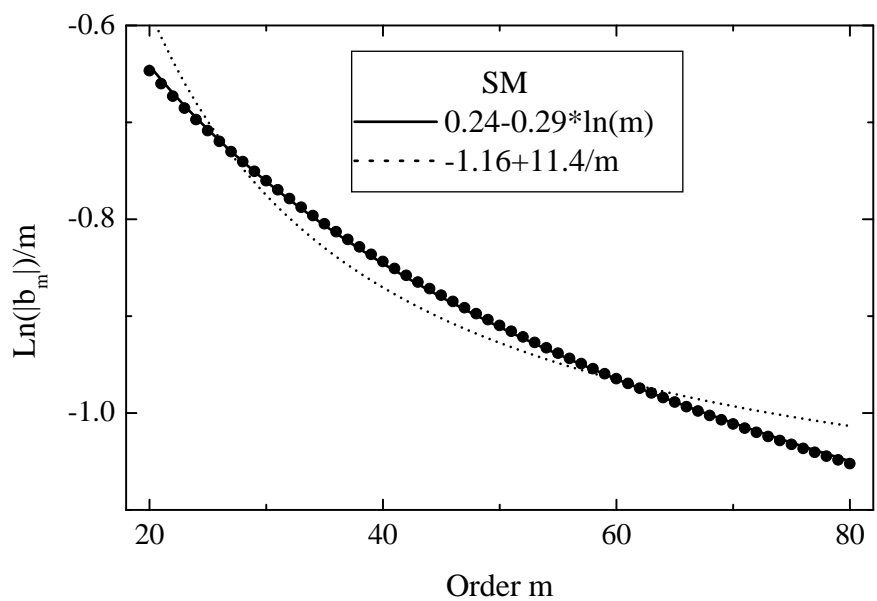,width=5.0in}}
\vskip125pt
\caption{Comparison of fits of the form $A_1+A_2/m$ (dots)and
$B_1{\rm ln}(m)+B_2$ (solid line) 
with the data provided in Fig. \ref{fig:logco}
for the SM (circles).}
\label{fig:toyfit}
\end{figure}

For the inverse function $d_1(y_1)$,
the situations more complicated as shown in Fig. \ref{fig:logco}.
The quantity plotted in this figure will be used to discriminate
between a finite and an infinite radius of convergence. If 
$|b_m|\sim R^{-m}$ as for a radius of convergence $R$, then 
we have ${\rm ln}(|b_m|)/m \sim -{\rm ln}(R)+A/m$ for some constant $A$.
On the other hand, if $|b_m|\sim (m !)^{-\alpha}$ as for an infinite 
radius of convergence, then 
we have ${\rm ln}(|b_m|)/m \sim -\alpha({\rm ln}(m)-1)$.
In the following, we will compare fits of the form $A_1+A_2/m$ and
$B_1{\rm ln}(m)+B_2$.
We first consider the case of SM, where according to the our study in
Section \ref{sec:tm}, expect an infinite radius of 
convergence. This possibility is highly favored as shown in Fig. 
\ref{fig:toyfit}.
One sees clearly that the solid line is a much better fit.
The chi-square for the solid line fit is 200 times smaller.
In addition $B_1\simeq-B_2$ as expected.

\begin{figure}
\centerline{\psfig{figure=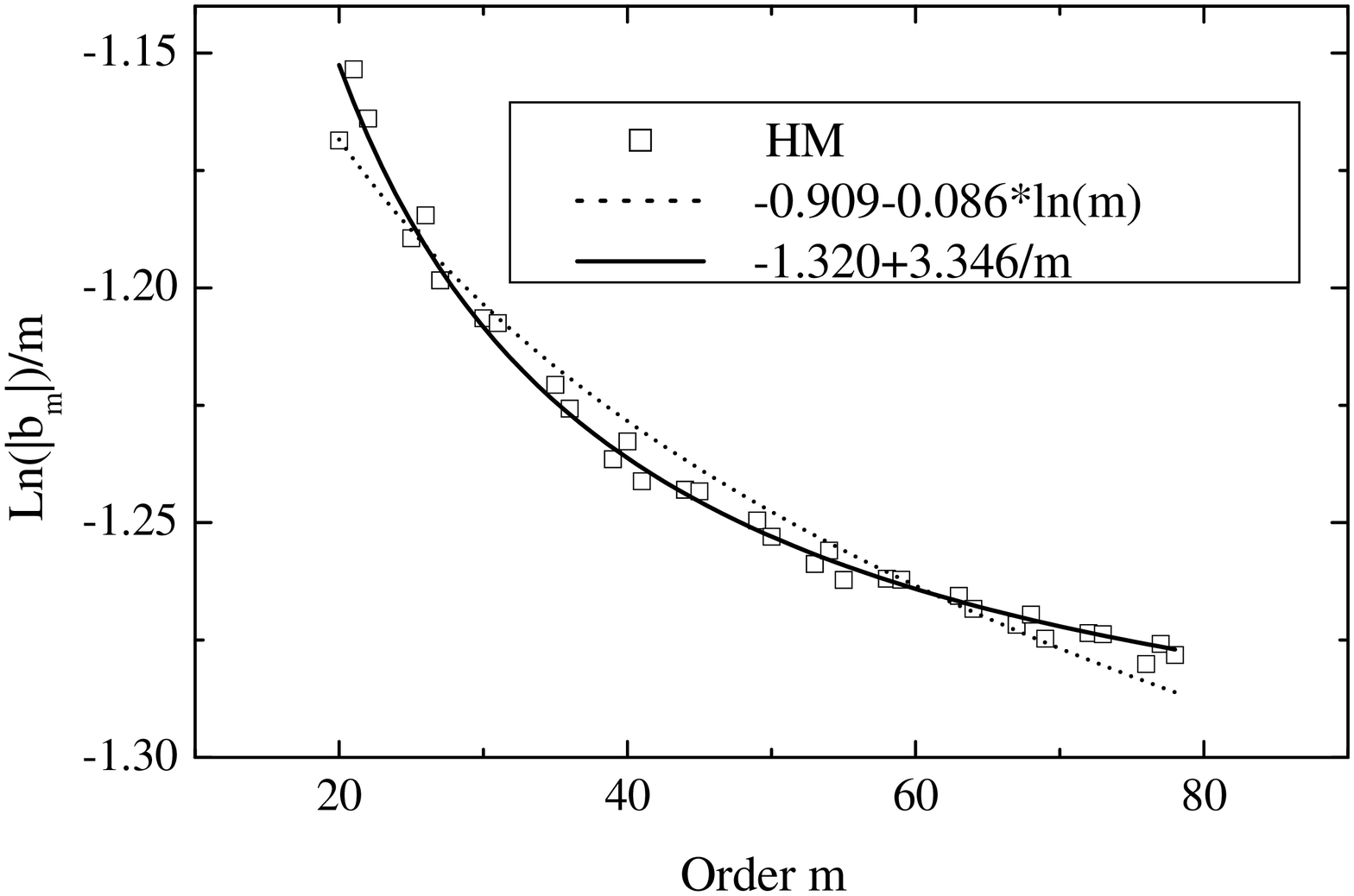,width=5.0in}}
\vskip125pt
\caption{Comparison of fits of the form $A_1+A_2/m$ (solid line)and
$B_1{\rm ln}(m)+B_2$ (dots) with selected points of the data 
in Fig. \ref{fig:logco}
for the HM  (boxes).}
\label{fig:hmfit}
\end{figure}

The analysis for the HM is more delicate. One observes periodic
``dips'' in Fig. \ref{fig:logco} which make the ratio analysis almost 
impossible. We have thus only considered, the ``upper envelope'' by 
removing the dips from Fig. \ref{fig:logco}. The fit represents an upper 
bound rather than the actual coefficients. The fits of the upper envelope
are shown in Fig. \ref{fig:hmfit}.
The possibility of a finite radius of convergence is slightly favored, the 
chi-square being 0.4 of the one for the other possibility. Also, the second 
fit does not have the $B_1\simeq-B_2$ property. From $A_1\simeq-1.32$, we 
estimate that the radius of convergence is about 3.7. 
This means that if we want to write some analytical formula for the 
flows by first calculating the initial values of the scaling fields
and then calculating ${\bf d}$ at successive iterations by using their
expression in terms of the scaling fields, we will have to switch 
variables in the crossover region.

\begin{figure}
\centerline{\psfig{figure=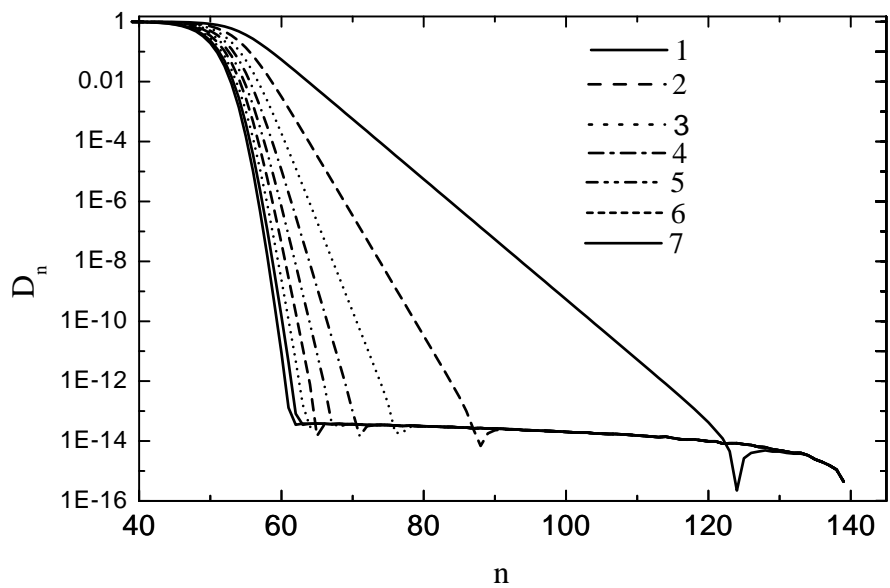,width=5.0in}}
\vskip125pt
\caption{The quantity $D_n$ defined in Eq. (\ref{eq:bdn}) for expansions 
of $\tilde{y}_1$ in $\beta$, at order 1 (solid line), 2 (dashes), 3 (dots),
etc... for each iteration $n$.}
\label{fig:app}
\end{figure}

As explained above, the approach of the HT fixed 
point is intrinsically a multivariable problem. For this reason, the
calculation of $D_n$, defined in Eq. (\ref{eq:bdn}), that tests the 
quality of scaling, will be our 
main tool of analysis.
In the following, we 
limit the discussion to $\tilde{y}_1({\bf h})$ which enters in the 
susceptibility. We 
have calculated 
$\tilde{y}_1({\bf h})$ in terms of 11 variables,  
up to order 11 in $\beta$. 
As in section \ref{sec:tm}, we will use an empirical series
$a_{n,l}$, calculate the corresponding $h_{n,l}$ an plug them 
in the scaling fields and check the scaling properties. 
This empirical series was calculated with an initial Ising measure and 
$\beta=\beta_c -10^{-8}$ (see Ref. \cite{gam3}).
Again we define a quantity $D_n$ as in Eq.
(\ref{eq:bdn}) which is very small when we have good scaling and increases
when the approximation breaks down. The results are shown in 
Fig. \ref{fig:app} for successive orders in $\beta$.
The solid line on the right is the linear approximation. It becomes optimal
near $n=130$. The next line (dashes) is the second order in $\beta$ expansion.
It becomes optimal near $n=90$. Each next order gets 
closer and closer to be optimal near $n=60$. 
The last curve on the left is the order 9 approximation. It is hard to 
resolve the next two approximations on this graph.

The asymptotic value is stabilized with 16 digits and one may wonder
why we get only scaling with 14 or 15 digits in Fig. \ref{fig:app}.
The reason is that we use empirical data and that numerical errors can 
add coherently as explained in Ref. \cite{numerr}. This can be seen 
directly by considering the difference between two successive values 
of $D_n$. A detailed analysis shows 
that the numerical errors at each step tend to be negative 
more often than positive, and consequently there is a small ``drift''
which affects the last digits.

\begin{figure}
\centerline{\psfig{figure=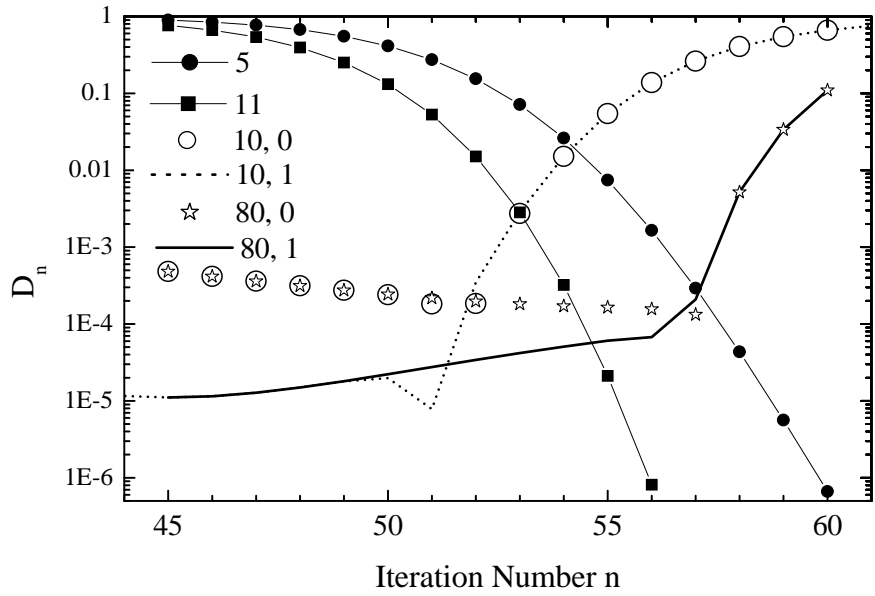,width=5.0in}}
\vskip125pt
\caption{Values of $D_n$ for $\tilde{y}_1$ up to order 5 
in $\beta$ (filled circle)
and 11 (filled boxes), and for $y_1$ up to order 10 in $d_1$ (empty circles)
and with first order corrections in $y_2$ (dots), and up to order 80
in $d_1$ (empty stars) and with first order corrections in $y_2$ (solid line).}
\label{fig:hmoverlap}
\end{figure}

We can now look at the $D_n$ defined as in Eq. (\ref{eq:bdn}) 
for $y_1$ and $\tilde{y}_1$ together 
in Fig. \ref{fig:hmoverlap}.
If we use an expansion of order 5 in $\beta$ for $\tilde{y}$ 
and of order 10 in $d_1$ for $y_1$, we can get scaling within a few 
percent for both variables at $n=54$. We can go below 1 part in 1000,
with an expansion of order 11 in $\beta$ and order 80 in $d_1$.
At this point, the main problem is that the effects of the subleading 
correction makes the scaling properties {\it worse} when $n\leq 57$
and $n$ decreases. One can improve the scaling properties by taking
the effects of $y_2$ into account. A detailed study shows that one 
can estimate the subleading effects between $n=40$ and $n=45$ as
\begin{equation}
\frac{y_1(d_{n,1})}{\lambda_1^n}\simeq 7.2778\times 10^{-9}+3.2
\times 10^{-9}\times\lambda_2^n 
\label{eq:cosca}
\end{equation}
It is thus possible to get a function scaling better by subtracting these
correction. This improve the scaling properties by almost one order of 
magnitude near $n=54$ and by almost two order of magnitude near $n=45$.
From Fig. \ref{fig:hmoverlap}, we see that the combined scaling
is optimal near $n=54$ which corresponds to an approximate value
of 2 for $y_1$.

Using Eq. (\ref{eq:pertexp}), we can 
take into account the first order correction in $y_2$. 
After simple manipulations, we can rewrite the RG invariant as 
\begin{equation}
\Theta \simeq C_1+C_2y_{2,0}(y_{1,0})^{\Delta}  \ ,
\label{eq:rginv}
\end{equation}
with $C_1=G(y_1)y_1^{\gamma}$ and 
$H(y_1)y_1^{\gamma-\Delta}$\ . The functions $G$ and $H$ are defined by
Eq. (\ref{eq:paramy}). They rely on {\it both} expansions used and 
consequently they are only valid in a crossover region.
Using explicit forms of $C_1$ and $C_2$ as a function of $y_1$, we 
observe a plateau for each function which are shown in Fig. \ref{fig:rginv}.
Using the flattest part of the plateau to estimate the constant we
obtain $C_1\simeq 1.46416$ and $C_2\simeq1.663$. Note that these 
two numbers are dependent of the choice of the scales for the 
scaling fields, but independent of the initial conditions. Consequently, 
if everybody agreed on the scales, these quantities could be called 
universal.
\subsection{step 3}

There remains to estimate the initial values of the $y_1$ and $y_2$.
This step will done numerically from empirical 
values of ${\bf d}_n$. First, we obtain a 
rough estimate of $y_{2,n}$ from $d_{2,n}$ at values of $n$ 
where the linear approximation is good. Plugging this value in 
Eq. (\ref{eq:pertexp}) for $l=1$, and 
inverting to get $y_{n,1}$ as a function of $d_{n,1}$. Dividing by 
$\lambda_1^n$ we get an estimated value of $y_{0,1}$ with a 
plateau of about 
10 iterations where the value is stabilized with 6 digits. Repeating for 
various temperatures we were able to determine the leading 
coefficient in Eq. (\ref{eq:init1}):
$Y_{1;1}\simeq 0.72782$.
Using  Eq. (\ref{eq:pertexp}) but for $d_2$, together with our 
previous estimates of $y_{1;n}$, we obtain
$Y_{2;0}\simeq -0.565$. Subleading coefficients are difficult to resolve
because $2\Delta\sim1$ and the nonlinear contributions in $y_{2}$ are 
of the same order as the analytical corrections.

This conclude our approximate treatment of step 3. 
Using Eqs. (\ref{eq:suspar}) and (\ref{eq:rginv}) we obtain the usual 
parametrization of the susceptibility of Eq. (\ref{eq:param}) with  
$A_0\simeq 2.1162$ and $A_1\simeq -1.196$  in very good agreement with
a fully numerical determination of these amplitudes \cite{gam3}.
 
\section{conclusions}

In summary, we have shown with two examples that the 3 steps advocated in 
Section \ref{sec:3steps} lead to an accurate determination of the leading and 
subleading critical amplitudes. We provided numerical evidence 
that the formal expansions of the scaling fields associated with the  
HT and unstable fixed point have reasonable convergence properties and 
scale properly in overlapping domains. The determination of the initial
values of the scaling fields associated with the unstable fixed point 
(third step) have been 
obtained numerically in the case of an initial Ising measure. 
Putting everything together, we were able to 
confirm numerical results obtained earlier.

Analytical methods are now being used \cite{lilipro} 
to consider initial measures of the 
Landau-Ginzburg type
in the vicinity of the Gaussian fixed point. For such initial measures, 
we can use perturbation theory 
in the quartic (or higher orders) 
coupling constant to construct the scaling fields associated 
with the Gaussian fixed point.
Interestingly, one could use some of the methods developed here to 
interpolate between the Gaussian fixed point and the unstable 
fixed point. The completion of this task will allow us 
to give analytical expressions 
for the renormalized quantities in terms of the bare ones, which is 
the notoriously difficult problem that has to be faced by a field theorist.
We are planning to extend the calculations 
performed here, to higher order derivatives of the free energy with a 
non-zero magnetic field and 
check explicitly amplitudes relations  
appearing in the literature
\cite{aharony83,aharony80,chang80}. 
Another interesting question that could be addressed within this context 
is the crossover from classical to critical behavior 
\cite{luijten98,bagnuls85,pelissetto98}.
The completion of these projects will provide a 
detailed comparison between general RG expectations and their 
practical realization for the HM.

The hierarchical approximation used in this article has 
allowed us to calculate large order expansions for the scaling fields.
The fact that the general ideas advocated have worked properly means
that one should now attempt to apply them to 
nearest neighbor models where similar calculations
would be more time consuming. The examples we have in mind  
are spin models in three dimensions and asymptotically free
theory such as the $O(N)$ spin models in two dimensions or lattice
gauge theory in four dimensions. 
In all cases, there exists some advanced technology for 
the weak and strong coupling expansions but the question of the interpolation
has only been studied with the Monte Carlo method \cite{gonzalez87,taro00}.

We expect that some of the simple features found in the study of the 
HT fixed point of the HM will generalize to nearest neighbor models.
First, the fact that the HT scaling fields are linearly related to 
the successive derivatives of the free energy. Second, the fact
that the restriction to a finite number of eigenvectors of the linearized
RG transformation near the HT fixed point can be used to obtain improved 
HT expansions such as the polynomial truncation used above. 
However, the most difficult task remains a construction of the 
unstable fixed point and the RG flows in its vicinity, with a control 
comparable to the case the hierarchical model. 
This is a challenging problem for the future.

\begin{acknowledgments}
This research was supported in part by the Department of Energy
under Contract No. FG02-91ER40664.
Y. M. thanks the 
Aspen Center for Physics, where discussions
have motivated part of this work 
and where part of 
this manuscript was written.

\end{acknowledgments}

\begin{references}
\bibitem{privman}
V. Privman, P. Hohenberg and A. Aharony, in {\em Phase Transitions and Critical
Phenomena}, v. 14, C. Domb and J. Lebowitz Editors
(Academic Press, New York, 1991), and references therein.
\bibitem{poincare92}
H. Poincar\'e, {\em Les Methodes Nouvelles de la Mecanique Celeste}
  (Gauthier-Villars, Paris, 1892).
\bibitem{arnold88}
V. Arnold, {\em Geometrical Methods in the Theory of Ordinary Differential
  Equations} (Springer-Verlag, New York, 1988).


\bibitem{wegner72}
F. Wegner, Phys. Rev. B {\bf 3},  4529  (1972).

\bibitem{cardy96}
J. Cardy, {\em Scaling and Renormalization in Statistical Physics} (Cambridge
  University Press, Cambridge, 1996).

\bibitem{gonzalez87}
A. Gonzalez-Arroyo and M. Okawa, Phys. Rev. D {\bf 35},  672  (1987).

\bibitem{luijten98}
E. Luijten and K. Binder, Phys. Rev. E {\bf 58},  R4060  (1998).

\bibitem{taro00}
P. de~Forcrand~et al., Nucl. Phys. B {\bf 577},  263  (2000).

\bibitem{bagnuls85}
C. Bagnuls and C. Bervillier, Phys. Rev. {\bf B32},  7209  (1985).

\bibitem{pelissetto98}
A. Pelissetto, P. Rossi, and E. Vicari, Phys. Rev. {\bf E58},  7146  (1998) and
Nucl. Phys. {\bf B554},  552  (1999);
A. Pelissetto and E. Vicari, cond-mat/0012164;
S. Caracciolo {\it et~al.}, cond-mat/0105160.

\bibitem{gaite96}
J. Gaite and D. O'Connor, Phys. Rev. D {\bf 54},  5163  (1996).

\bibitem{campostrini01}
M. Campostrini, J. Stat. Phys. {\bf 103},  369  (2001).

\bibitem{osc1}
Y. Meurice, G. Ordaz, and V.~G.~J. Rodgers, Phys.\ Rev.\ Lett. {\bf 75},  4555
  (1995).

\bibitem{osc2}
Y. Meurice, S. Niermann, and G. Ordaz, J.\ Stat.\ Phys.\ {\bf 87},  363
  (1997).

\bibitem{wilson71b}
K. Wilson, Phys. Rev. B. {\bf 4},  3185  (1971).

\bibitem{dyson69}
F. Dyson, Comm.\ Math.\ Phys.\ {\bf 12},  91  (1969).

\bibitem{baker72}
G. Baker, Phys.\ Rev.\ B {\bf 5},  2622  (1972).

\bibitem{dual}
Y. Meurice and S. Niermann, Phys. Rev. E {\bf 60},  2612  (1999).


\bibitem{nickel80}
B. Nickel,  in {\em Phase Transitions, Cargese 1980}, edited by M. Levy, J.~L.
  Guillou, and J. Zinn-Justin (Plenum Press, New York, 1982).


\bibitem{bleher75}
P. Bleher and Y. Sinai, Comm. Math. Phys. {\bf 45},  247  (1975).

\bibitem{collet78}
P. Collet and J. P. Eckmann, {\em A Renormalization Group Analysis of the
  Hierarchical Model in Statistical Mechanics} (Springer-Verlag, Berlin, 1978).

\bibitem{kw88}
H. Koch and P. Wittwer,  in {\em Nonlinear Evolution and Chaotic Phenomena},
Nato ASI Series, Series B: Physics, Vol. 176,
G. Gallavotti and P. Zweifel Editors, Plenum; and in
{\em Mathematical Quantum Field Theory and Related Topics}, 
Canadian Mathematical Society, Conference Proceedings v. 9 (1988).

\bibitem{finite}
J. Godina, Y. Meurice, M. Oktay, and S. Niermann, Phys. Rev. D {\bf 57},  6326
  (1998).

\bibitem{gam3}
J. Godina, Y. Meurice, and M. Oktay, Phys. Rev. D  {\bf 57},  R6581  (1998)
and {\bf 59},  096002  (1999).

\bibitem{koch95}
H. Koch and P. Wittwer, Math. Phys. Electr. Jour. {\bf 1},  Paper 6  (1995).


\bibitem{smalld}
Y. Meurice, Phys. Rev. E {\bf 63},  055101(R)  (2001).
\bibitem{ew87}
J. P. Eckmann and P. Wittwer, Jour. Stat. Phys. {\bf 46}, 455 (1987).

\bibitem{niemeijer76}
T. Niemeijer and J. van Leeuwen,  in {\em Phase Transitions and Critical
  Phenomena, vol. 6}, edited by C. Domb and M. Green (Academic Press, New York,
  1976).

\bibitem{lilipro}
L. Li and Y. Meurice, in preparation.
\bibitem{hyper}
J.~J. Godina, Y. Meurice, and M. Oktay, Phys. Rev. D {\bf 61},  114509  (2000).


\bibitem{numerr}
Y. Meurice and B. Oktay, Phys. Rev. D {\bf 63},  016005  (2001).


\bibitem{aharony83}
A. Aharony and M. Fisher, Phys. Rev. B {\bf 27},  4394  (1983).

\bibitem{aharony80}
A. Aharony and G. Ahlers, Phys. Rev. Lett. {\bf 44},  782  (1980).

\bibitem{chang80}
M. Chang and A. Houghton, Phys. Rev. Lett. {\bf 44},  785  (1980).



\end{references}

\vfill 
\eject
\vfill 
\eject
$  $
\vskip125pt
\vfill 
\eject
$ $
$ $
\vskip125pt

\vfill 
\eject
$ $
\vskip125pt
\vfill 
\eject
$ $
\vskip125pt
\vfill 
\eject
$ $
\vskip125pt
\vfill 
\eject
$ $
\vskip125pt
\vfill 
\eject
$ $
\vskip125pt
\vfill 
\eject
$ $
\vskip125pt
\vfill 
\eject
$ $
\vskip125pt
\vfill 
\eject
$ $
\vskip125pt
\noindent
\vfill 
\eject
$ $
\vskip125pt
\vfill 
\eject
$ $
\vskip125pt
$ $
\vskip125pt

\end{document}